\documentclass[conf]{new-aiaa}
\usepackage[utf8]{inputenc}

\usepackage{graphicx}
\usepackage{amsmath}
\usepackage[version=4]{mhchem}
\usepackage{siunitx}
\usepackage{longtable,tabularx}
\usepackage{float}
\usepackage{subfig}
\usepackage{placeins} 
\usepackage{xcolor}

\title{Study on the Resolution of Large-Eddy Simulations for Supersonic Jet Flows}

\author{Diego F. Abreu\footnote{Ph.D. Candidate, Graduate Program in Space Sciences 
and Technologies, Departamento de Ciência e Tecnologia Aeroespacial, DCTA/ITA; 
E-mail: mecabreu@yahoo.com.br.}}
\affil{Instituto Tecnológico de Aeronáutica, 12228--900, São José dos Campos, SP, Brazil}
\author{Carlos Junqueira-Junior\footnote{Research Engineer, Arts et Métiers Institute of 
Technology, DynFluid laboratory; E-mail: junior.junqueira@ensam.eu.}}
\affil{Arts et Métiers Institute of Technology, DynFluid, CNAM, HESAM University, 
151 Boulevard de l'Hôpital, 75013, Paris, France}
\author{Eron T. V. Dauricio\footnote{Ph.D. Candidate, Graduate Program in Space Sciences and 
Technologies, Departamento de Ciência e Tecnologia Aeroespacial, DCTA/ITA; 
E-mail: eron.tiago90@gmail.com.}}
\affil{Instituto Tecnológico de Aeronáutica, 12228--900, São José dos Campos, SP, Brazil}
\author{João Luiz F. Azevedo\footnote{Senior Research Engineer, Aerodynamics Division, 
Departamento de Ciência e Tecnologia Aeroespacial, DCTA/IAE/ALA; 
E-mail: joaoluiz.azevedo@gmail.com. Fellow AIAA.}}
\affil{Instituto de Aeronáutica e Espaço, 12228--904, São José dos Campos, SP, Brazil}

\begin{document}

\maketitle

\begin{abstract}
The present study is concerned with large-eddy simulations (LES) of supersonic jet flows. 
The work addresses, in particular, the simulation of a perfectly expanded free jet flow with an exit 
Mach number of 1.4 and an exit temperature equal to the ambient temperature. Calculations are 
performed using a nodal discontinuous Galerkin method. The present effort studies the effects of mesh and 
polynomial refinement on the solution. The present calculations consider computational meshes and 
plynomial orders such that the number of degrees of freedom (DOFs) in the solution ranges from 50 
to 410 million. Mean velocity results and root mean square (RMS) values of velocity fluctuations 
indicate a better agreement with experimental data as the resolution is increased. 
The generated data provide a good understanding of the effects of increasing the discretization 
refinement for LES calculations of jet flows. The present results can guide future simulations of similar 
flow configurations.
\end{abstract}

\section{Introduction}
With the progress of computing power in the last years, the large-eddy simulation (LES) formulation appears as an alternative to Reynolds-averaged Navier-Stokes (RANS) methods due to its reasonable cost when compared to the direct numerical simulation (DNS) of the Navier-Stokes equations or even physical experiments.  LES can provide valuable information on complex configurations such as shear layers \cite{Bres2019,Kumar2017} and detached flows \cite{Ghate2021,Masoudi2021} due to its capability to generate unsteady data for flow and temperature fields with high-frequency fluctuations, which are necessary for aerodynamics, acoustics, loads, and heat transfer analyses.

The authors are interested in the LES of jet flows from aircraft and rockets engines \cite{Junior2018,Junior2020,BogeyMarsden2016,Bres2017,Debonis2017}. More specifically, on the perfectly expanded configuration, when the jet exit pressure matches the ambient pressure, at 1.4 Mach number. 
%
%
%
Recent work highlights \cite{Abreu2021} the effects of structured second-order finite-difference and unstructured nodal discontinuous-Galerkin spatial discretizations \cite{Kopriva2010,Hindenlang2012} on the flow of interest at a fixed number of degrees of freedom (DOF). The results indicate good agreement with experimental and numerical data, where the spatial resolution is sufficient and with the same order of error in the coarser mesh regions. Therefore, the current study addresses the effects of refinement on the LES of a supersonic jet flow configuration using the FLEXI framework \cite{Krais2021}. The solver applies an unstructured nodal discontinuous Galerkin spatial discretization that allows evaluating the influence of mesh and polynomial ({\em hp}) refinement.


The literature \cite{BogeyBailly2010,Mendezetal2012,Bresetal2018,ShenMiller2019,Corriganetal2018} does not agree on the mesh requirements for adequately solving the LES formulation due to employing different numerical methods for solving jet flows. The present paper studies the effects of mesh and polynomial refinement along with mesh topology to identify the minimum mesh requirements for adequately solving the problem of interest. The research group improved the baseline grid from Ref.\@ \cite{Abreu2021} with local mesh refinement in the vicinity of the lipline and with an increase in the number of elements, ranging from $6.2 \times 10^6$ to $15.4 \times 10^6$ elements. The jet flow calculations use second-order and third-order polynomials. The simulations present 50 to 410 million DOFs when combining grid and polynomial refinement.


The generated data for mean velocity and RMS of velocity fluctuations are investigated and compared with experimental data \cite{BridgesWernet2008} at different regions of the domain where the jet is developing. 
The paper is organized to introduce the reader to the description of physical and numerical formulation in the second section. Then, one can find details of the experimental configuration and the numerical setup in sections three and four. Finally, the results and the concluding remarks close the work in sections five and six, respectively.

\section{Numerical Formulation}
\subsection{Governing Equations}

The work has the interest in the solution of the filtered Navier-Stokes equations. The filtering strategy is based on a spatial filtering process that separates the flow into a resolved part and a non resolved part. Usually the filter size is obtained from the mesh size. The filtered Navier-Stokes equations in conservative form can be written by

\begin{equation}
\frac{\partial \mathbf{\bar{Q}}}{\partial t} + \nabla \cdot \mathbf{F} ( \mathbf{\bar{Q}}, \nabla \mathbf{\bar{Q}})=0,
\label{eq.1}
\end{equation}
where $\mathbf{\bar{Q}}=[\bar{\rho}, \bar{\rho} \tilde{u}, \bar{\rho} \tilde{v}, \bar{\rho} \tilde{w}, \bar{\rho} \check{E}]^{T}$ is the vector of filtered conserved variables and $\mathbf{F}$ is the flux vector. The flux vector can be divided into the Euler fluxes and the viscous flux, $\mathbf{F}=\mathbf{F}^e-\mathbf{F}^v$. The fluxes with the filtered variables may be written as
\begin{equation}
\mathbf{F}_i^e= \left[ \begin{array}{c} 
                    \bar{\rho} \tilde{u}_i \\ \bar{\rho} \tilde{u} \tilde{u}_i + \delta_{1i} \bar{p}\\ \bar{\rho} \tilde{v} \tilde{u}_i + \delta_{2i}\bar{p} \\ \bar{\rho} \tilde{w} \tilde{u}_i + \delta_{3i}\bar{p} \\ (\bar{\rho} \check{E} + \bar{p}) \tilde{u}_i
                   \end{array} \right]  \hspace*{1.5 cm} 
\mathbf{F}_i^v= \left[ \begin{array}{c} 
                    0 \\ \tau_{1i}^{mod} \\ \tau_{2i}^{mod} \\ \tau_{3i}^{mod} \\ \tilde{u}_j \tau_{ij}^{mod} - q_i^{mod}
                   \end{array} \right] \hspace*{1.5 cm} \mbox{ , for } i = 1,2,3 ,                  
\end{equation}
where $\tilde{u}_i$ or $(\tilde{u}, \tilde{v}, \tilde{w})$ are the Favre averaged velocity components, $\bar{\rho}$ is the filtered density, $\bar{p}$ is the filtered pressure and $\bar{\rho} \check{E}$ is the filtered total energy per unit volume. The terms $\tau_{ij}^{mod}$ and $q_{i}^{mod}$ are the modified viscous stress tensor and heat flux vector, respectively, and $\delta_{ij}$ is the Kronecker delta. The filtered total energy per unit volume, according to the definition proposed by \citet{Vreman1995} in its "system I", is given by
\begin{equation}
\bar{\rho} \check{E} = \frac{\bar{p}}{\gamma - 1} + \frac{1}{2}\bar{\rho}\tilde{u}_i\tilde{u}_i.
\end{equation}

The filtered pressure, Favre averaged temperature and filtered density are correlated using the ideal gas equation of state $\bar{p}= \bar{\rho} R \tilde{T}$, and $R$ is the gas constant, written as $R = c_p - c_v$. The properties $c_p$ and $c_v$ are the specific heat at constant pressure and volume, respectively. The modified viscous stress tensor may be written as
\begin{equation}
\tau_{ij}^{mod}=(\mu + \mu_{SGS}) \left(\frac{\partial \tilde{u}_i}{\partial x_j} + \frac{\partial \tilde{u}_j}{\partial x_i} \right) - \frac{2}{3} (\mu + \mu_{SGS}) \left(\frac{\partial \tilde{u}_k}{\partial x_k} \right) \delta_{ij} 
\end{equation}
where $\mu$ is the dynamic viscosity coefficient, calculated by Sutherland's Law, and $\mu_{SGS}$ is the SGS dynamic viscosity coefficient, which is provided by the subgrid-scale model. The strategy of modeling the subgrid-scale contribution as an additional dynamic viscosity coefficient is based on the Boussinesq hyphotesis. The modified heat flux vector, using the same modeling strategy, is given by
\begin{equation}
q_i^{mod}=-(k+k_{SGS})\frac{\partial \tilde{T}}{\partial x_i}
\end{equation}
where $k$ is the thermal conductivity coefficient of the fluid and $k_{SGS}$ is the SGS thermal conductivity coefficient given by
\begin{equation}
k_{SGS}=\frac{\mu_{SGS} c_p}{Pr_{SGS}}
\end{equation}
and $Pr_{SGS}$ is the SGS Prandtl number.
%
%
The present work employs only the static Smagorinsky model \cite{Smagorinsky1963} to calculate the subgrid-scale contribution.

\subsection{Nodal Discontinuous Galerkin Method}

The nodal discontinuous Galerkin method used in this work is based on the modeling proposed by \citet{Kopriva2010}, and \citet{Hindenlang2012}. In this discretization, the domain is divided into multiples hexahedral elements. This choice of elements permits that the interpolating polynomial be defined as a tensor product basis with degree $N$ in each space direction. The implementation is simpler and improve the computational efficiency of the code. 

In this method, the elements from the physical domain are mapped onto a reference unit cube elements $E=[-1,1]^3$. The equations, presented in Eq.\@ (\ref{eq.1}) need also to be mapped to this new reference domain, leading to

\begin{equation}
J \frac{\partial \mathbf{\bar{Q}}}{\partial t} + \nabla_{\xi} \cdot \bar{\mathcal{F}} = 0,
\label{eq.2}
\end{equation}
where $\nabla_{\xi}$ is the divergence operator with respect to the reference element coordinates, $\mathbf{\xi}=(\xi^1,\xi^2,\xi^3)^T$, $J= \arrowvert \partial \mathbf{x} / \partial \mathbf{\xi} \arrowvert$ is the Jacobian of the coordinate transformation and $\bar{\mathcal{F}}$ is the contravariant flux vector.

The discontinuous Galerkin formulation is obtained multiplying Eq.\@ (\ref{eq.2}) by the test function $\psi=\psi(\xi)$ and integrating over the reference element $E$
\begin{equation}
\int_E J \frac{\partial \mathbf{\bar{Q}}}{\partial t} \psi d \xi + \int_E \nabla_{\xi} \cdot \bar{\mathcal{F}} \psi d \xi = 0.
\label{eq.3}
\end{equation}
It is possible to obtain the weak form of the scheme by integrating by parts the second term in Eq.\@ (\ref{eq.3})
\begin{equation}
\frac{\partial}{\partial t} \int_E J \mathbf{\bar{Q}} \psi d \xi + \int_{\partial E} (\bar{\mathcal{F}} \cdot \vec{N})^* \psi dS - \int_E \bar{\mathcal{F}} \cdot (\nabla_{\xi} \psi ) d \xi = 0,
\label{eq.4}
\end{equation}
where $\vec{N}$ is the unit normal vector of the reference element faces. Because the discontinuous Galerkin scheme allows discontinuities in the interfaces, the surface integral above is ill-defined. In this case, a numerical flux, $\bar{\mathcal{F}}^*$, is defined, and a Riemann solver is used to compute the value of this flux based on the discontinuous solutions given by the elements sharing the interface.

For the nodal form of the discontinuous Galerkin formulation, the solution in each element is approximated by a polynomial interpolation of the form
\begin{equation}
\mathbf{\bar{Q}}(\xi) \approx \sum_{p,q,r=0}^N \mathbf{\bar{Q}}_h(\xi_p^1,\xi_q^2,\xi_r^3,t)\phi_{pqr}(\xi),
\end{equation}
where $\mathbf{\bar{Q}}_h(\xi_p^1,\xi_q^2,\xi_r^3,t)$ is the value of the vector of conserved variables at each interpolation node in the reference element and $\phi_{pqr}(\xi)$ is the interpolating polynomial. For hexahedral elements, the interpolating polynomial is a tensor product basis with degree N in each space direction
\begin{equation}
\phi_{pqr}(\xi)=l_p(\xi^1)l_q(\xi^2)l_r(\xi^3), \hspace{10pt} l_p(\xi^1)= \prod_{\substack{i=0 \\ i \ne p}}^{N_p} \frac{\xi^1-\xi_i^1}{\xi_p^1-\xi_i^1}.
\end{equation}
The definitions presented are applicable to other two directions.

The numerical scheme used in the simulation additionally presents the split formulation presented by \citet{Pirozzoli2011}, with the discrete form given by \citet{Gassner2016}, to enhance the stability of the simulation. The split formulation is employed to Euler fluxes only. The solution and the fluxes are interpolated and integrated at the nodes of a Gauss-Lobatto Legende quadrature, which presents the summation-by parts property, that is necessary to employ the split formulation.

The Riemann solver used in the simulations is a Roe scheme with entropy fix \cite{Harten1983} to ensure that second law of thermodynamics is respected, even with the split formulation. To be able to adequately handle the viscous flux in the boundaries of the elements, the lifting scheme of \citet{BassiRebay1997} is used, which is also known as BR2. The time marching method chosen is the five-stage, fourth-order explicit Runge-Kutta scheme of \citet{CarpenterKennedy1994}. 

The shock waves that appear in the simulation are stabilized using the finite-volume sub-cell shock capturing method of \citet{Sonntag2017}. Even though the methodology used in the simulation solves the discontinuous Galerkin approach, it only handle discontinuities in the interface of the elements. The shock capturing method permits to stabilize the simulation with shock waves inside the elements.

\section{Experimental Configuration}

%
The focus of this work is to investigate the influence of mesh and polynomial refinement on the perfectly 
expanded jet flow, which is present in many applications, such as supersonic military aircraft and large 
launch vehicles. The experimental work of \citet{BridgesWernet2008} provides data flow properties for 
different jet flow configurations
%
%
In this work, the interest is to simulate the fully expanded free jet flow configuration with a Mach number 
of $1.4$. In this configuration the jet flow has a static pressure in the nozzle exit plane that equals the 
ambient static pressure with a supersonic velocity. For such a flow configuration, the shock waves are 
weaker when compared to other operating conditions, which reduces the constraints of mesh refinement and,
consequently, the computational cost of the simulation.

The experimental apparatus for the analysed configuration is composed of a convergent-divergent nozzle 
designed based on the method of characteristics \cite{BridgesWernet2008}. The nozzle exit diameter is 
$50.8$ mm. The Reynolds number based on nozzle exit diameter is 
$1.58 \times 10^6$.
%
%
The experimental data acquisition applies the Time-Resolved Particle Image Velocimetry (TRPIV) at a $10$ kHZ sample rate. The investigation uses two sets of cameras, one captures the flow along the nozzle centerline, and the other captures the flow of the mixing layer along the nozzle lipline.


\section{Numerical Setup}

\subsection{Geometry and Mesh Configuration}

\begin{figure}[htb!]
\centering
	\includegraphics[width=0.65\linewidth]{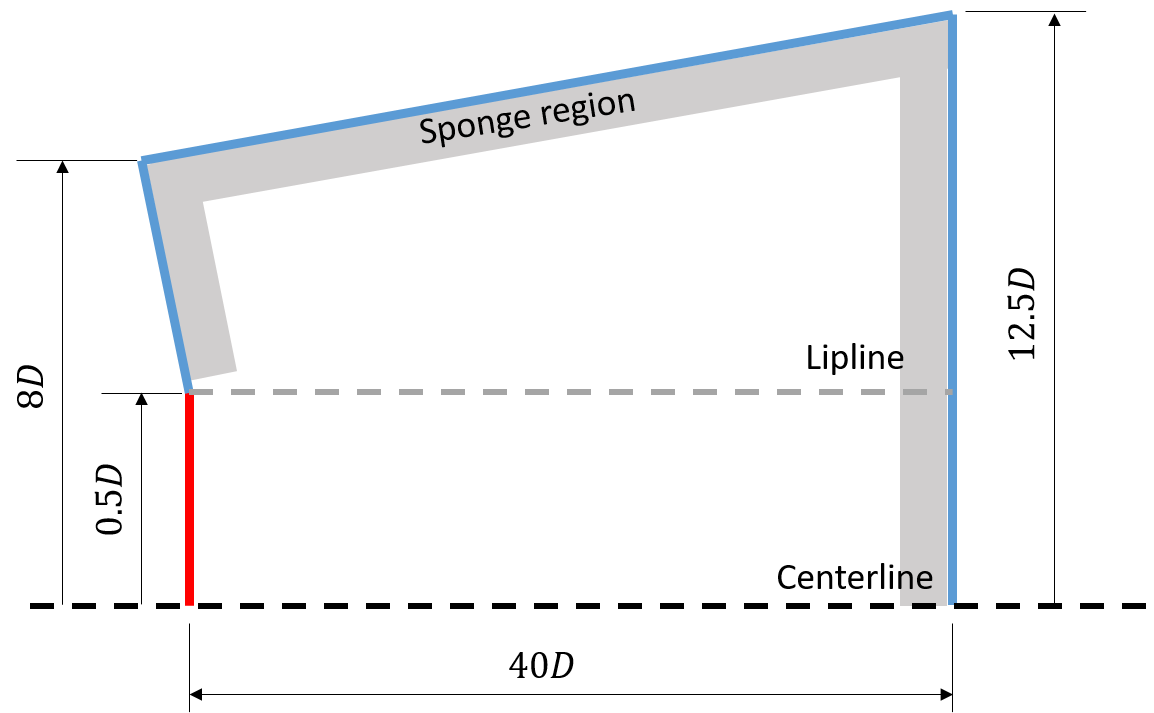}
    \caption{2-D schematic representation of the computational domain used on the
	jet flow simulations.}
    \label{fig.geo}
\end{figure}

The geometry used for the calculations in this work presents a divergent shape and axis 
length of $40D$, where $D$ is the jet inlet diameter and has external diameters of $16D$ and 
$25D$. Figure \ref{fig.geo} illustrates a 2-D representation of the computational domain,
indicating the inlet surface in red, the farfield region in blue, the lipline in grey, and
the centerline in black. 
Two different computational grids are used in the present work. The coarser mesh used here, named 
M-1 mesh, is the same grid developed in Ref.\ \cite{Abreu2021}. The other computational grid, which 
is termed M-2 mesh here, was developed specifically for the present effort and it represents a 
considerable improvement over the M-1 mesh. The modifications in the M-2 mesh are both topological 
and the result of an increase in the number of grid cells. These modifications result in a much 
higher refinement level around the jet inlet, encompassing both the original jet as well as the 
strong mixing region around the jet. Afterwards, this mesh transitions to a uniform grid point 
distribution as one moves downstream in the longitudinal direction. The mesh generation uses a 
multiblock strategy in order to handle hexahedral cells.
%

The grid design attempts to capture the jet flow until a distance of $x/D=15$ from the 
inlet surface, indicated in red in Fig.\ \ref{fig.geo}. Then, the size of elements 
increases as an attempt to dissipate frequencies that could destabilize the simulation. 
Previous results \cite{Junior2020} support the creation of mesh topology, indicating 
that a surface with one degree of opening angle would better represent the middle 
surface of the jet mixing layer. From this surface, two regions rise to increase 
the resolution of the domain. They have a geometrical stretching in the section $x/D=0$ 
to enable local refinement in the shear region of the flow and a uniform distribution 
in $x/D = 15$. The internal section of the mixing layer connects to one hexahedral block 
that forms the core of the mesh. That hexahedral block has its dimensions defined to keep 
the size of the elements equal to the size of the last cells in the region of the mixing 
layer.

\begin{figure}[htb!]
\centering
\subfloat[Radial mesh refinement ($\Delta y/D$) in the longitudinal sections $x/D=0$ and $x/D=15$.]{
	\includegraphics[width=0.47\linewidth]{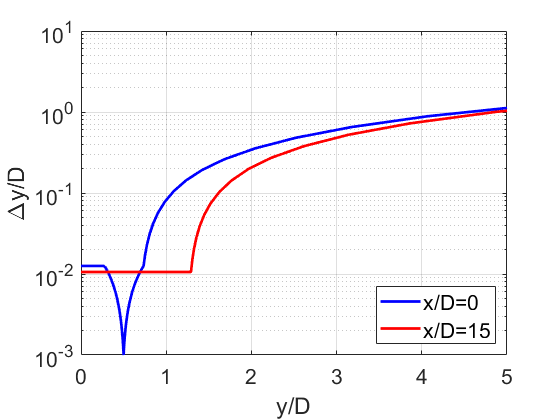}
	\label{fig.elem1}
	}
\subfloat[Longitudinal mesh refinement ($\Delta x/D$) along the jet centerline.]{
	\includegraphics[width=0.47\linewidth]{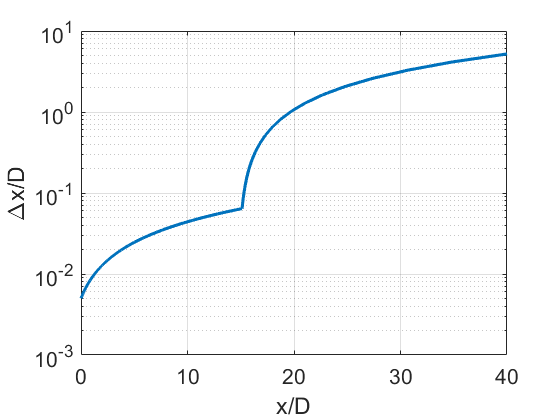}
	\label{fig.elem2}
	}
\caption{Distribution of grid spacing indicating radial and longitudinal refinement for the M-2 mesh.}
\end{figure}

\begin{figure}[htb!]
\centering
\subfloat[M-1 mesh.]{
	\includegraphics[width=0.47\linewidth]{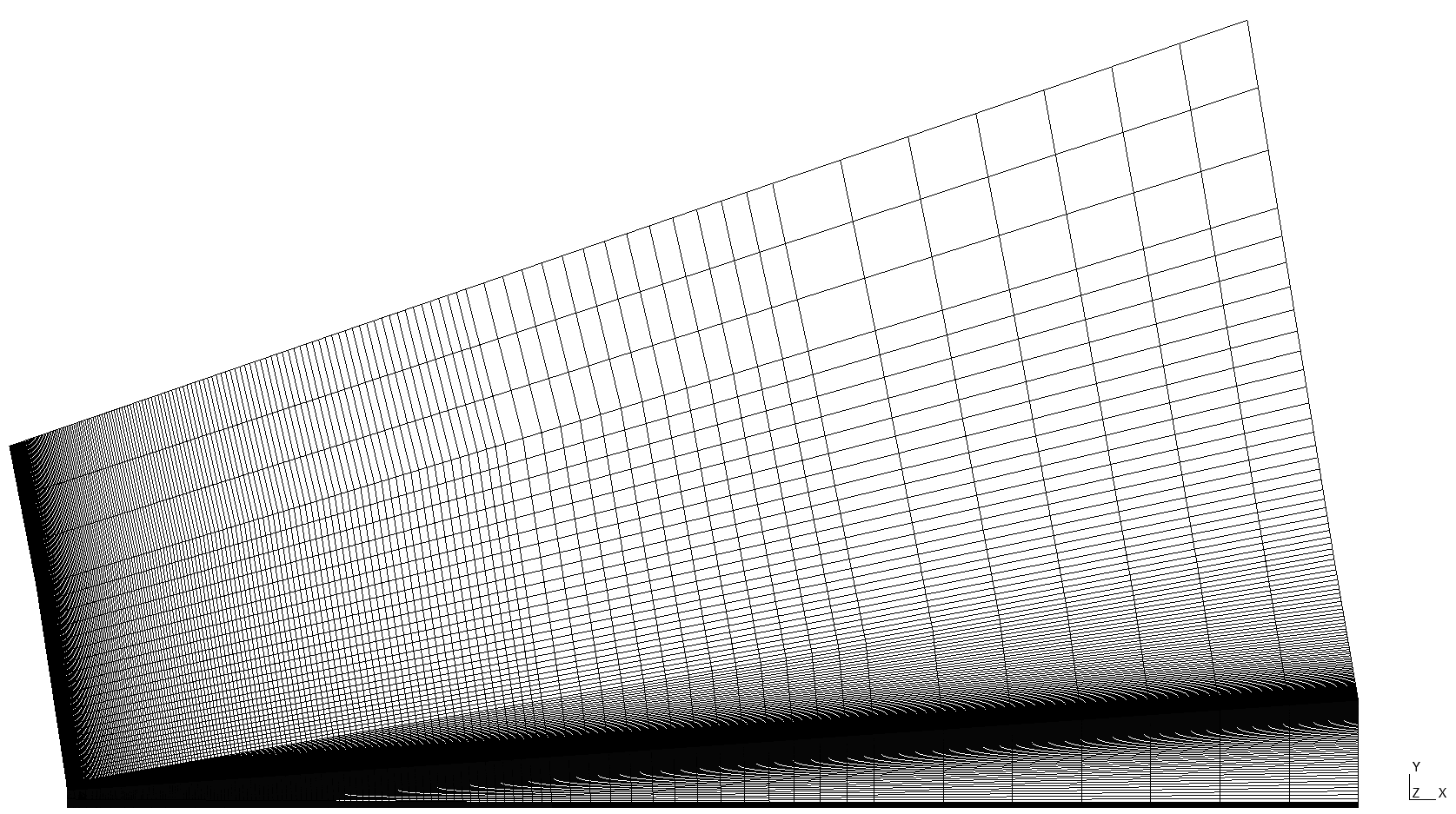}
	\label{fig.mes1}
	}
\subfloat[M-2 mesh.]{
	\includegraphics[width=0.47\linewidth]{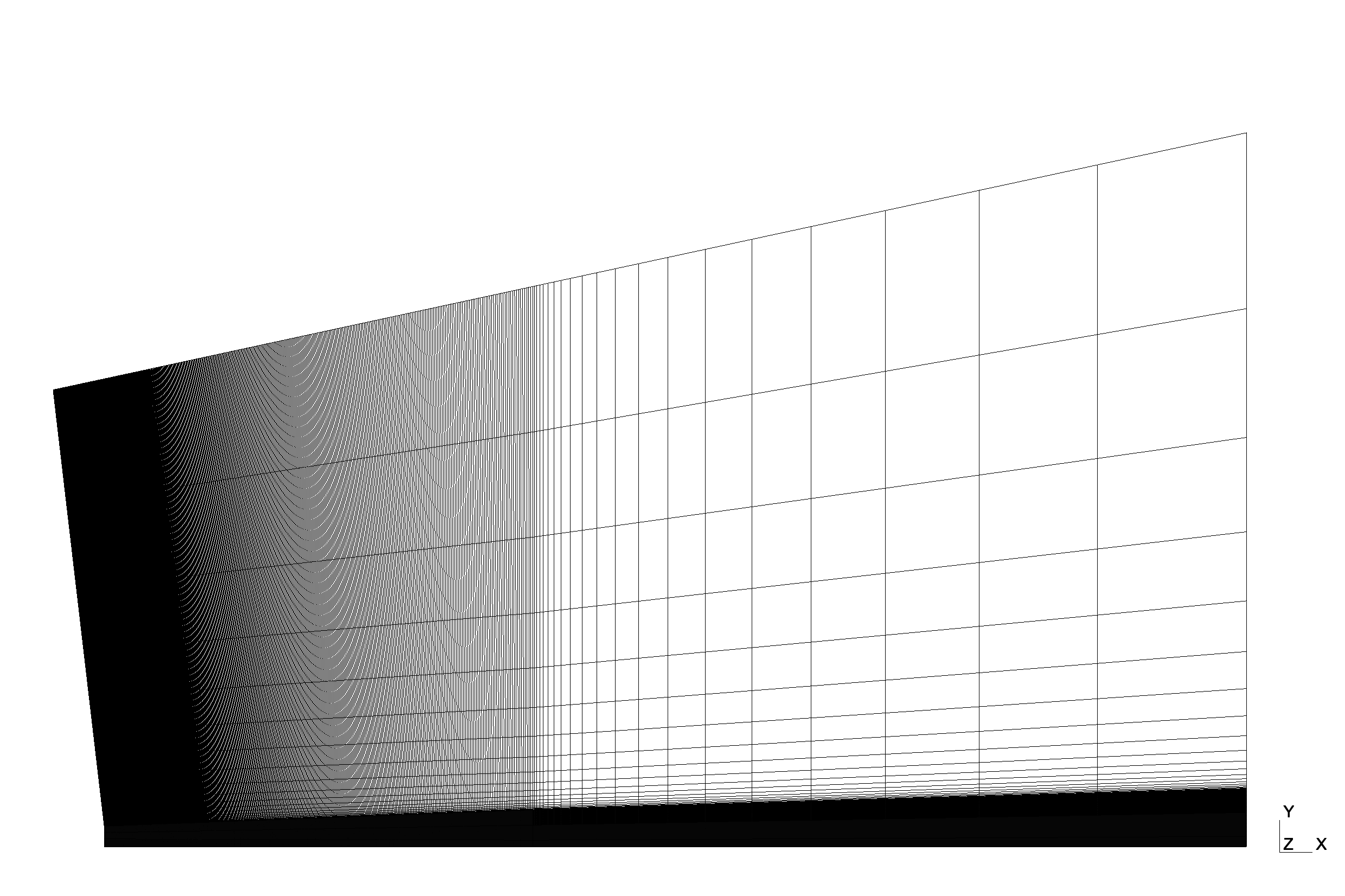}
	\label{fig.mesh2}
	}
\caption{Visualization of the half-plane longitudinal cutplanes for the meshes used in the present work.}
\label{fig.mesh}
\end{figure}

The grid refinement in the mixing layer is defined based on the literature 
\cite{BogeyBailly2010, BogeyMarsden2016, Bres2017, Debonis2017, Junior2018}. The grid 
spacing in the radial and axial directions along the mixing layer is $\Delta y_0/D=0.001$ 
and $\Delta x_0/D=0.005$, respectively. The centerline presents 651 elements set in 
geometrical stretching distribution between $x/D=0$ and $x/D=15$. Each region of the mixing layer 
contains 50 cells, and the Azimuthal direction accommodates 180 elements evenly 
distributed. Figure \ref{fig.elem1} presents the radial mesh 
refinement in two different longitudinal positions, $x/D=0$ and $x/D=15$. Figure 
\ref{fig.elem2} illustrates the axial mesh refinement along the jet centerline and 
Fig.\ \ref{fig.mesh} exhibits a cutplane of the mesh generated for the current paper and the baseline line mesh used in previous work.
The M-1 and M-2 grids have a total of 6.2 and 15.4 million cells, respectively,
and they are created with the GMSH \cite{Geuzaine2009} mesh generator.

\subsection{Boundary Conditions}

%
Properties on the jet inflow, $(\cdot)_{jet}$, and farfield, $(\cdot)_{ff}$, surfaces are 
indicated in Fig.\ \ref{fig.geo} in red and blue, respectively. 
A weakly enforced solution of a Riemann problem with a Dirichlet condition is 
enforced at the boundaries. The flow is characterized as perfectly expanded and 
isothermal, {\em i.e.} $p_{jet}/p_{ff}=T_{jet}/T_{ff}=1$, where $p$ stands for pressure and $T$ 
for temperature. The Mach number of the jet at the inlet is $M_{jet}=1.4$ and the Reynolds number 
based on the diameter of the nozzle is $Re_{jet} = 1.58 \times 10^6$. A small velocity component with 
$M_{ff}=0.01$ in the streamwise direction is imposed at the farfield to avoid numerical issues. A 
sponge zone \cite{Flad2014} is employed around the farfield boundaries, the gray area presented in 
Fig.\ \ref{fig.geo}, to damp any oscillations that could be reflected back to the jet.

\subsection{Simulation Settings and DOFs}

The current work compares the effects of {\it hp} refinement using three different calculations: 
S1, S2, and S3. The first simulation uses the M-1 computational grid, while the other two 
computations apply the M-2 mesh. Both studies, S1 and S2, employ first-order polynomial reconstructions 
in order to achieve second-order accuracy in spatial discretization. Calculation S3 uses second-order 
polynomial reconstructions in order to achieve a third-order accurate spatial discretization. The 
simulations, therefore, consider from 50 to 410 million DOFs. Table \ref{tab.mesh} indicates the 
settings used the three numerical studies performed in the present effort.
\begin{table}[htb!]
\centering
\caption{Summary of simulations settings.}
\begin{tabular}{ c  c  c  c  c  c } \hline \hline
Simulation & Meshes & Order of  & DOF/cell & Cells  & Total \# of DOF \\ 
 & & Accuracy & & ($10^{6}$) & ($10^{6}$) \\\hline
S1 & M-1 & 2nd order & 8 & $6.2$ & $\approx 50$ \\
S2 & M-2 & 2nd order & 8 & $15.4$ & $\approx 120$ \\ 
S3 & M-2 & 3rd order & 27 & $15.4$ & $\approx 410$ \\ \hline \hline
\end{tabular}
\label{tab.mesh}
\end{table}

\subsection{Calculation of Statistical Properties}

\label{chap.stats}

The simulation procedure involves three steps. The first one is to clean off the domain since 
the computation starts with a static flow initial condition. The simulations run three
flow-through times (FTT) to develop the jet flow. One FTT is the time required for one
particle with the jet velocity to cross the computational domain. In the sequence, the
simulations run an additional three FTT to produce a statistically steady condition. Then, 
in the last step, data are collected for another three FTT to obtain the statistical
properties of the flow.

%
The procedure for developing S3 simulation is slightly different. The simulation is a 
restart from the finished S2 calculation. The numerical framework FLEXI allows using 
one solution with different order of accuracy as initialization. Once the second-order
solution was already available, its usage was a short come to initialize the S3 
simulation. Hence, the S3 numerical study runs 0.5 FFT to allow the solution to adapt from 
second-order accuracy to third-order accuracy. Then it runs an additional 2 FTT to collect
data. Different frequencies of data acquisition were employed in each simulation. The S1 
case applies $160$ kHz, while S2 and S3 cases use $205$ and $225$ kHz, respectively.

\begin{figure}[htb!]
\centering
\includegraphics[width=0.8\linewidth]{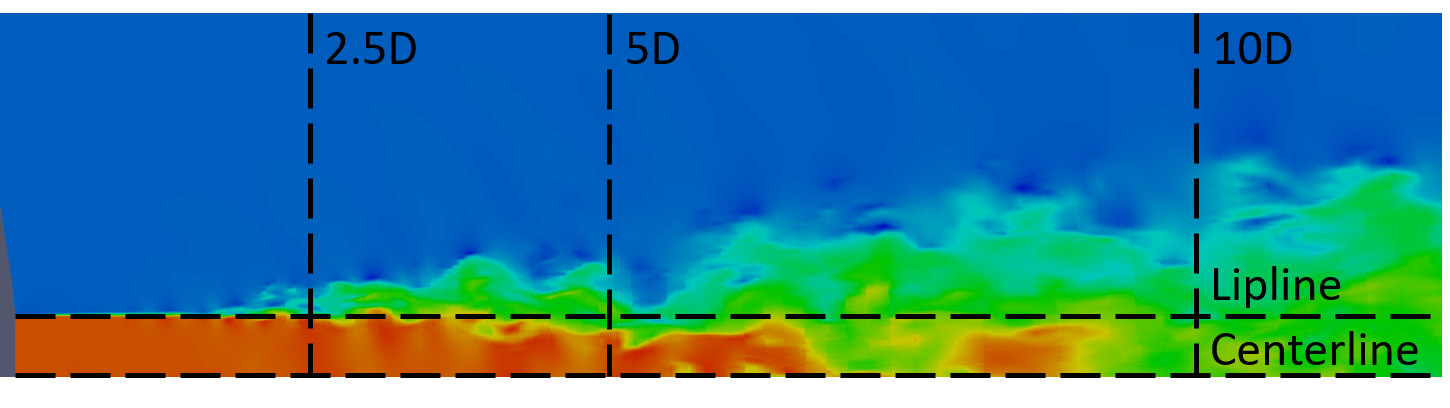}
\caption{Snapshot of the jet simulation with the two longitudinal lines and three crossflow lines 
along which data is extracted. Mach number contours are shown.}
\label{fig.jet_data_extract}
\end{figure}

The mean and the root mean square (RMS) fluctuations of properties of the flow are 
calculated along the centerline, lipline, and different domain surfaces in the 
streamwise direction. The centerline is defined as the line in the center of the 
geometry $y/D=0$, whereas the lipline is a surface parallel to the centerline and 
located at the nozzle diameter, $y/D=0.5$. The results from the lipline are an 
azimuthal mean from six equally spaced positions. The four surfaces in the streamwise 
positions are $x/D=2.5$, $x/D=5.0$, $x/D=10.0$, and $x/D=15.0$. Surface properties are
averaged using six equally spaced positions in the azimuthal direction. Figure
\ref{fig.jet_data_extract} illustrates a Mach number contours snapshot of the jet flow 
with the lines and surfaces of data extraction. 
%

\section{Results}

The results from S1, S2, and S3 simulations are presented in this section and compared 
to experimental data \cite{BridgesWernet2008}. The focus of this work is to assess the
resolution requirements for the correct prediction of supersonic jet flows. S1 and S2 
calculations are performed with the same polynomial order of accuracy, while S3 simulation 
uses third-order accuracy polynomials. The S1 Numerical study is performed with mesh M-1, 
with $6.2 \times 10^6$ elements, and S2 and S3 calculations are performed with mesh M-2, 
with $15.4 \times 10^6$ elements. The simulations have approximately $50$, $120$, and 
$410$ million DOFs.

\subsection{Velocity and Density Contours}

\begin{figure}[htb!]
\centering
\subfloat[S1 simulation.]{
	\includegraphics[trim = 0mm 30mm 0mm 50mm, clip, width=0.80\linewidth]{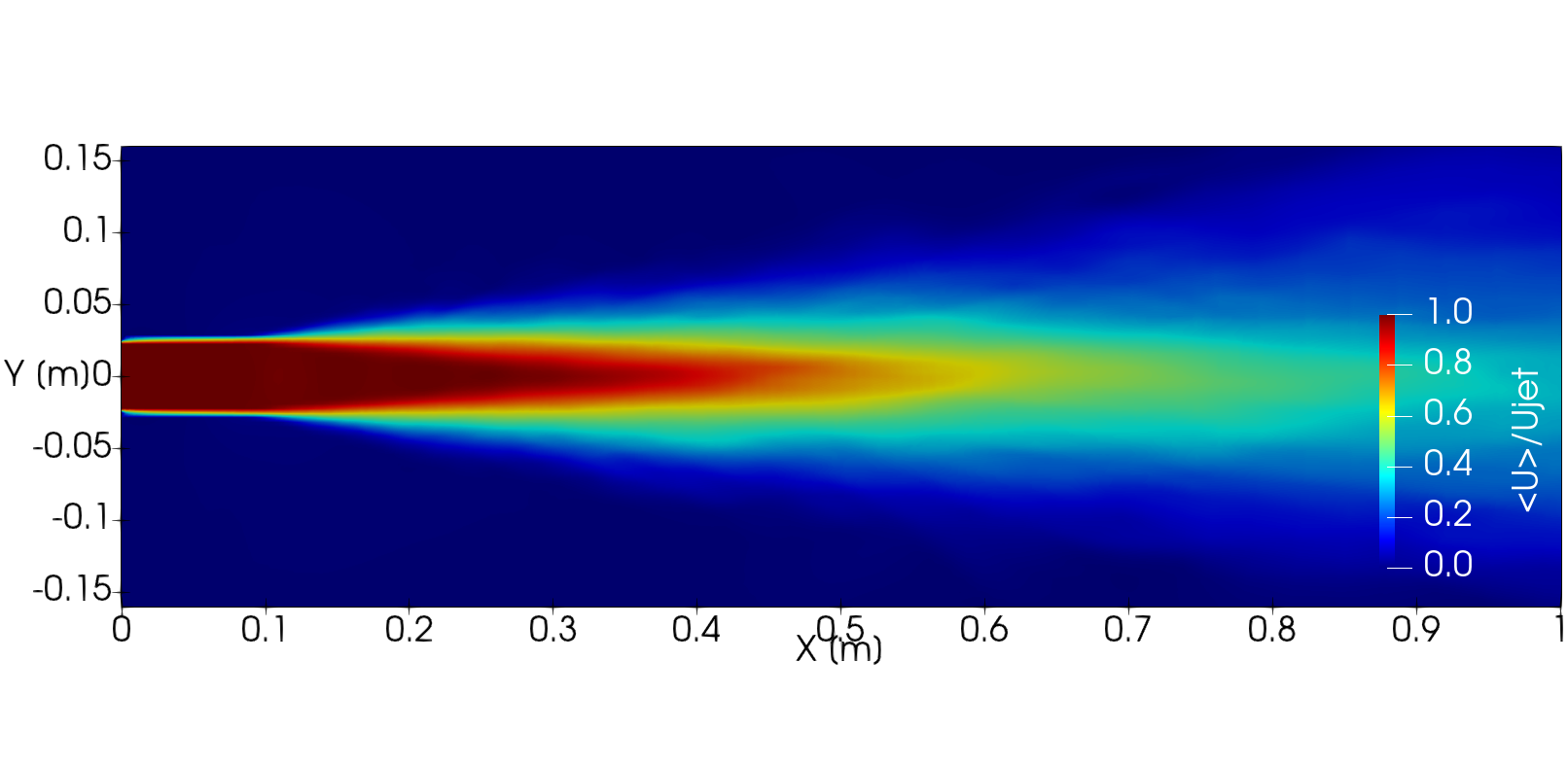}
	\label{fig.res_velxa}
	}
\newline
\subfloat[S2 simulation.]{
	\includegraphics[trim = 0mm 30mm 0mm 50mm, clip, width=0.80\linewidth]{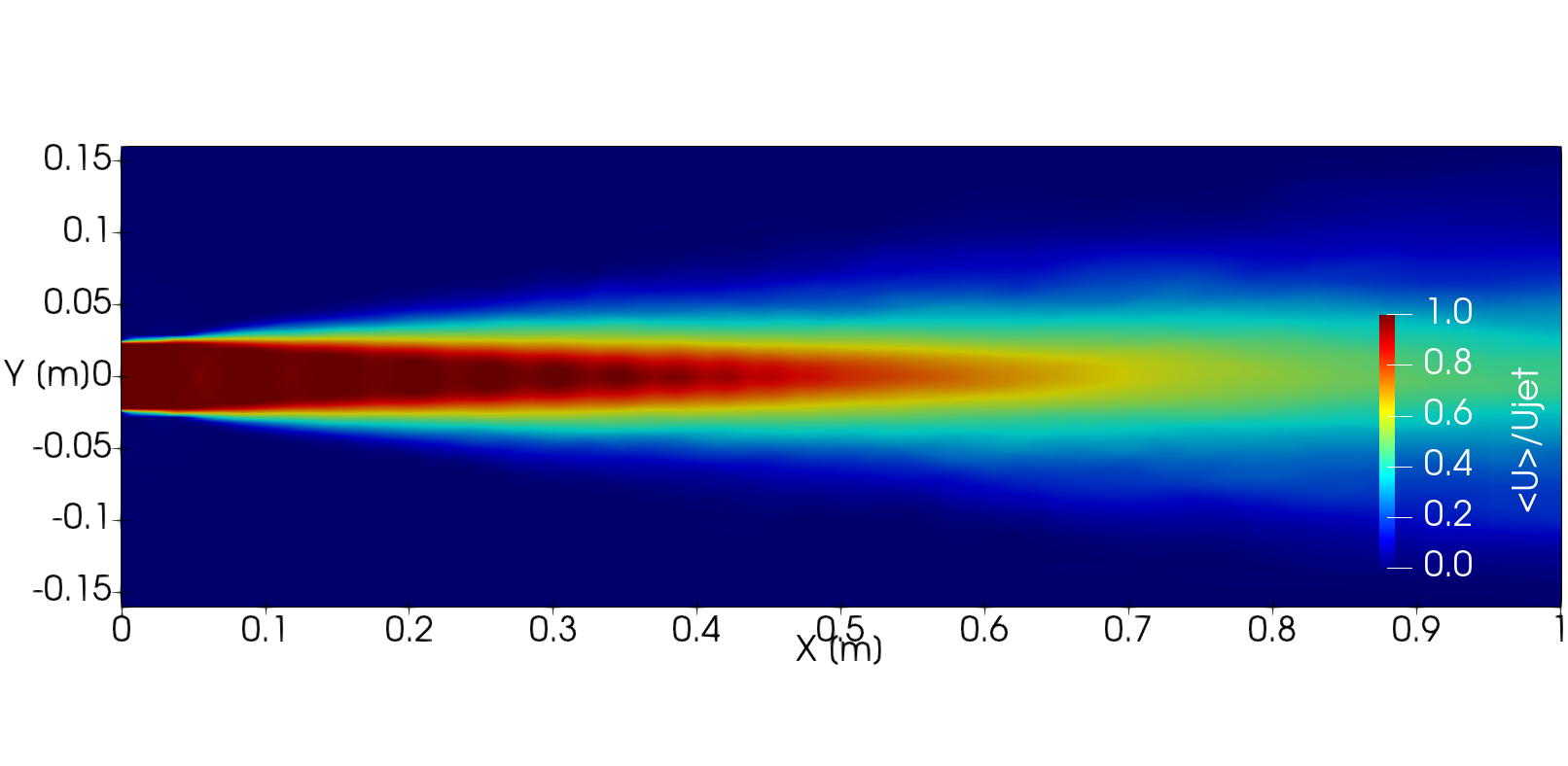}
	\label{fig.res_velxb}
	}
\newline
\subfloat[S3 simulation.]{
	\includegraphics[trim = 0mm 30mm 0mm 50mm, clip, width=0.80\linewidth]{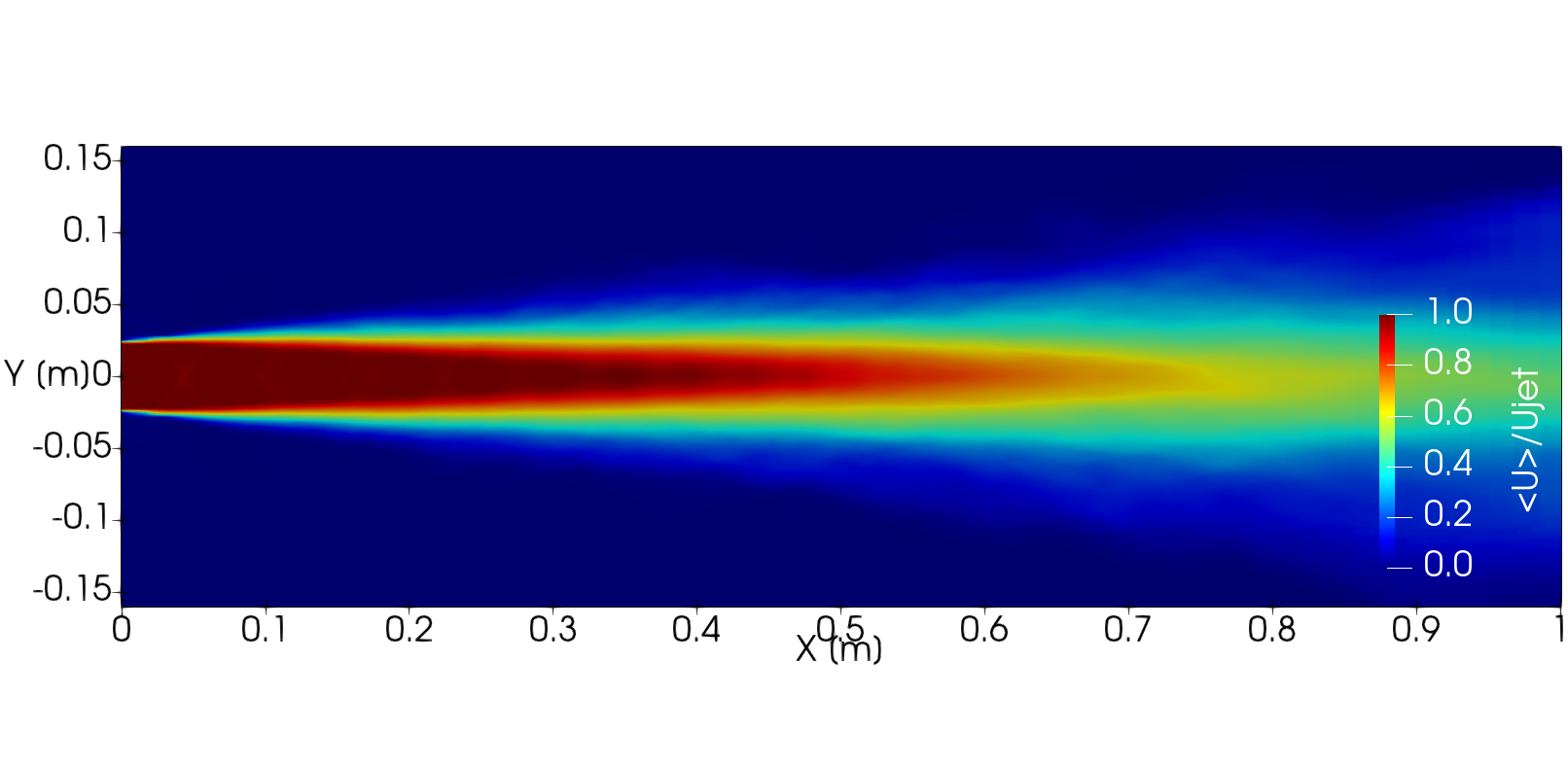}
	\label{fig.res_velxc}
	}
\newline
\caption{Contours of mean longitudinal velocity component along cutplanes in $z/D=0$ for the three simulations performed.}
\label{fig.res_velx}
\end{figure}

%
Initially, the contours of the mean longitudinal velocity component, RMS of longitudinal
velocity fluctuation, and mean density are presented for the three simulations. Figure
\ref{fig.res_velx} presents the contours of the mean longitudinal velocity component on 
a cut plane at $z/D=0$. The contours indicate qualitatively improvement in results when
increasing the number of DOFs. One can notice the size of the jet core is bigger in 
Fig.\ \ref{fig.res_velxc} than in Figs.\ \ref{fig.res_velxb} and \ref{fig.res_velxa}. 
Moreover, the development of the mixing layer start closer to the jet inlet in S3 
simulation than in S2 and S1 calculations. Furthermore, it is difficult to notice the
existence of shock waves in Fig.\ \ref{fig.res_velxa}, which are more clearly visible 
in Figs.\ \ref{fig.res_velxb} and \ref{fig.res_velxc}.

\begin{figure}[htb!]
\centering
\subfloat[S1 simulation.]{
	\includegraphics[trim = 0mm 30mm 0mm 50mm, clip, width=0.8\linewidth]{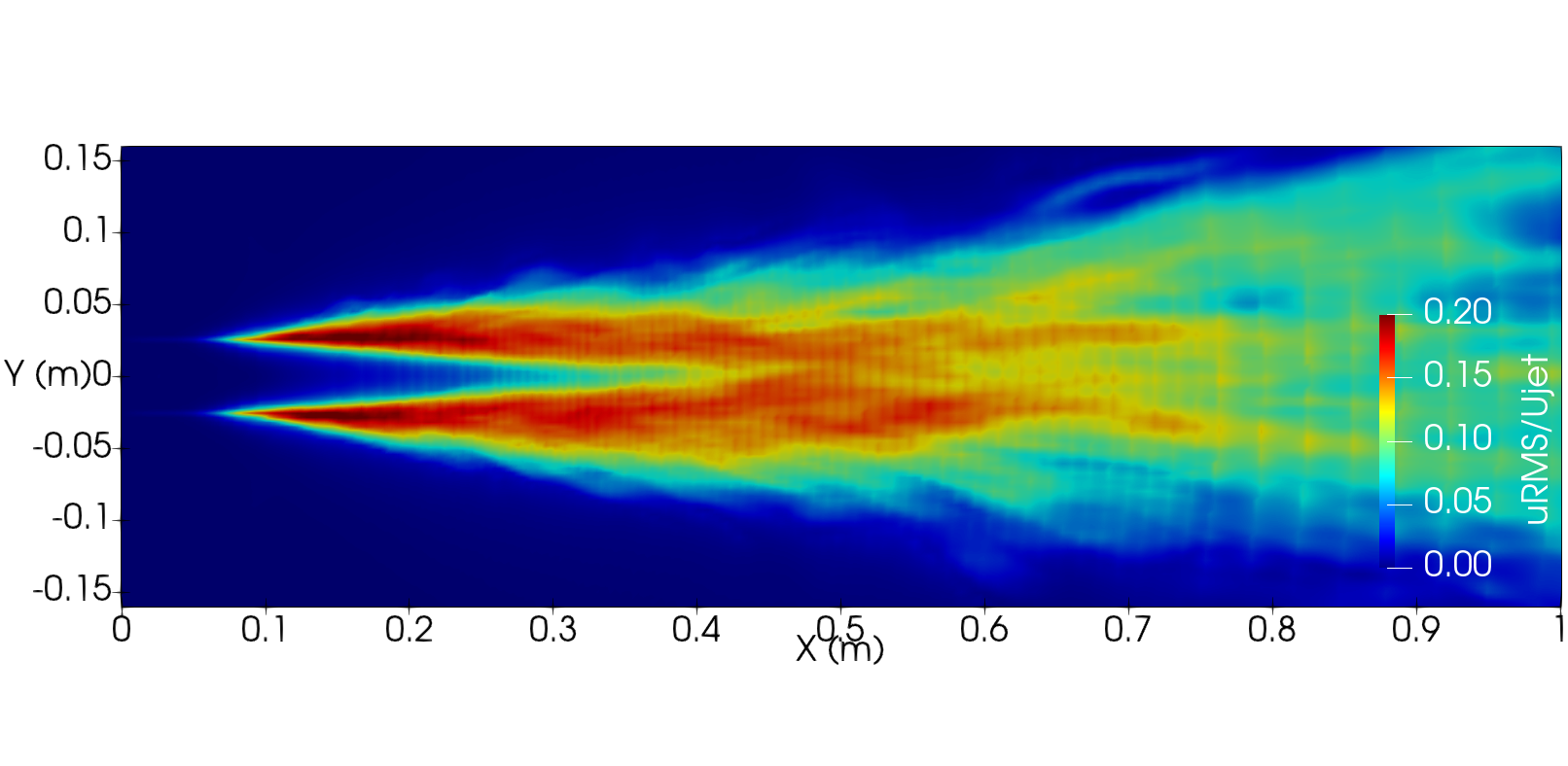}
	\label{fig.res_velxrmsa}
	}
\newline
\subfloat[S2 simulation.]{
	\includegraphics[trim = 0mm 30mm 0mm 50mm, clip, width=0.8\linewidth]{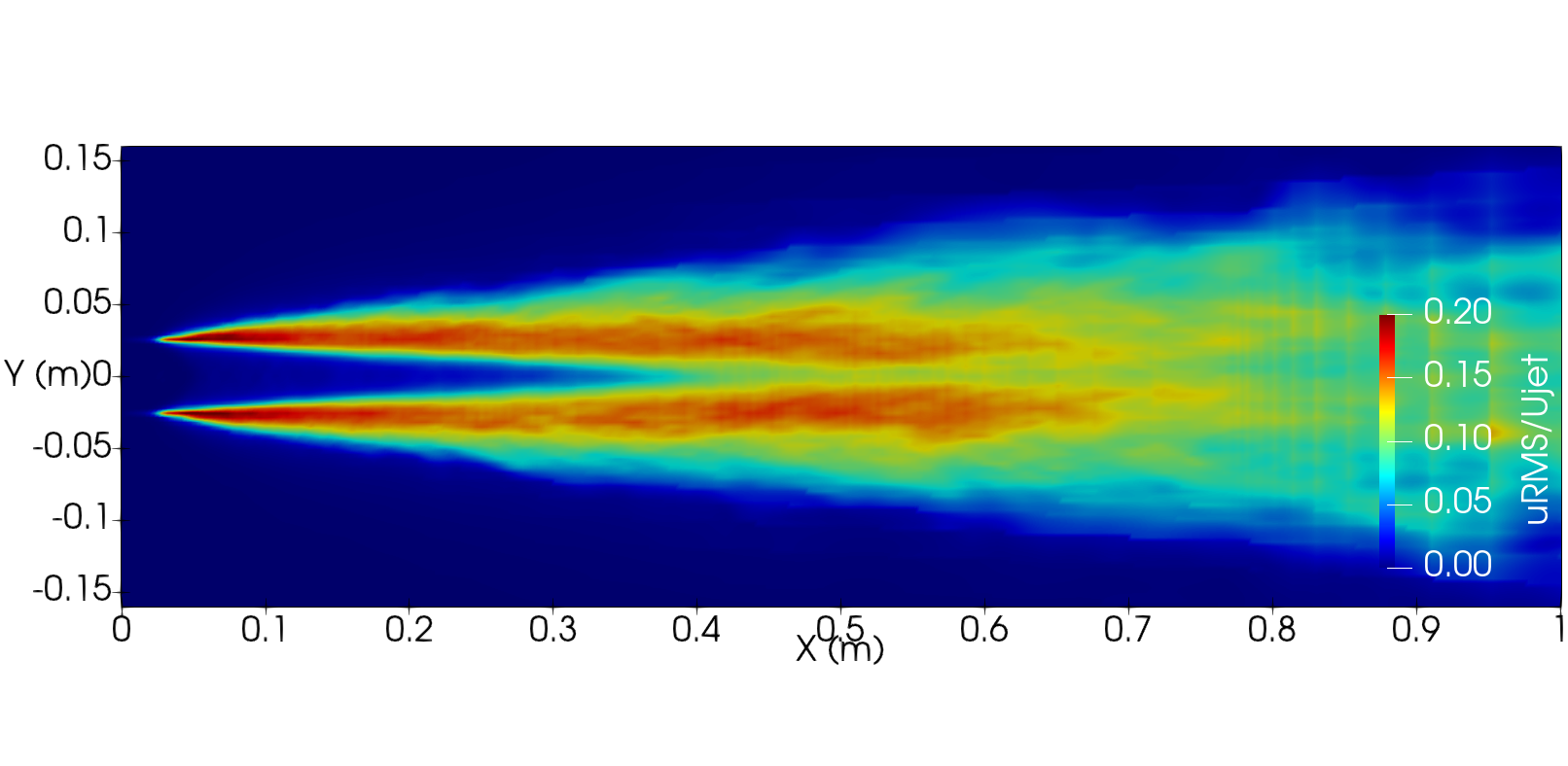}
	\label{fig.res_velxrmsb}
	}
\newline
\subfloat[S3 simulation.]{
	\includegraphics[trim = 0mm 30mm 0mm 50mm, clip, width=0.8\linewidth]{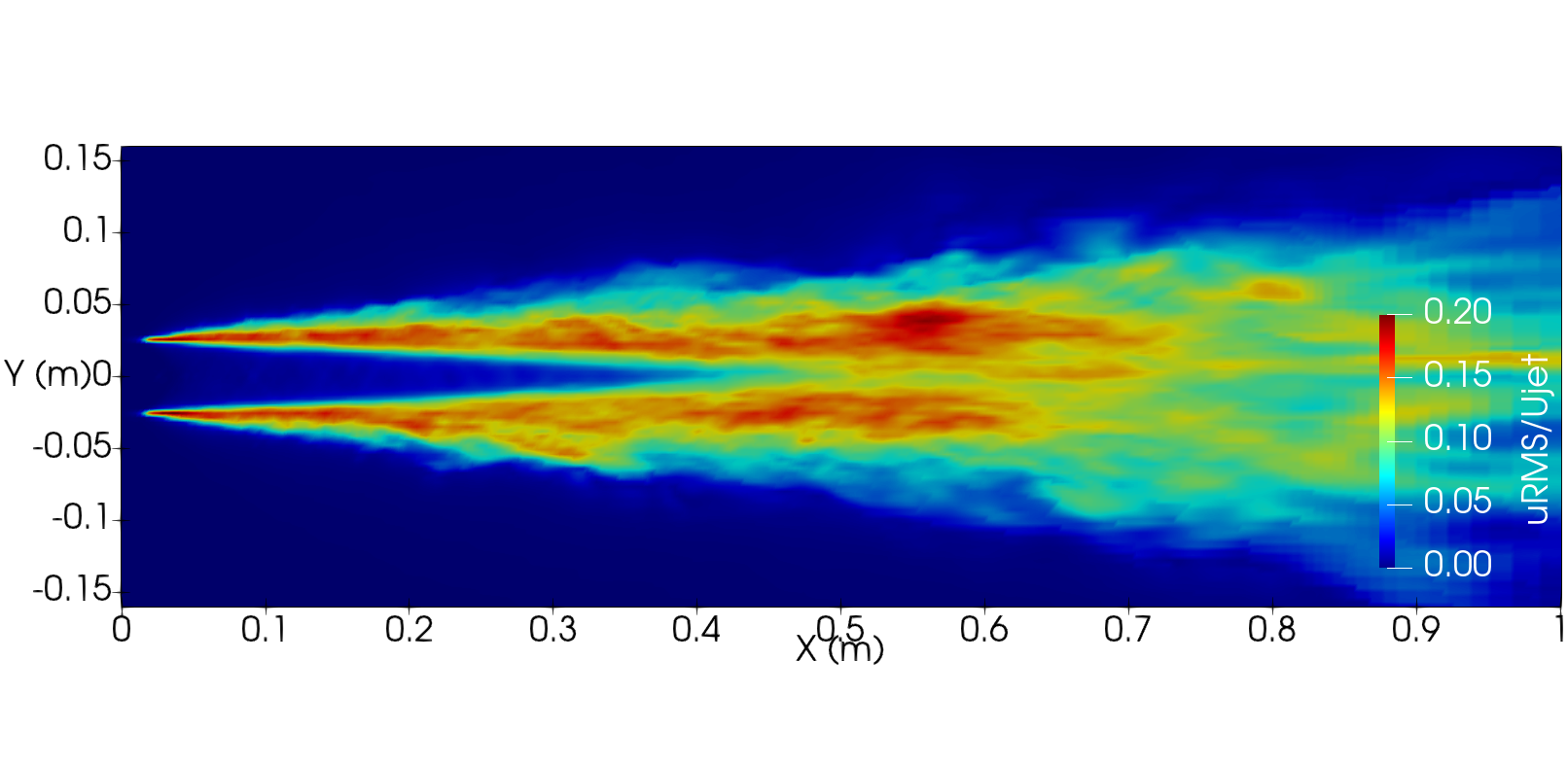}
	\label{fig.res_velxrmsc}
	}
\newline
\caption{Contours of RMS values of longitudinal velocity component fluctuations along cutplanes in $z/D=0$ for the three simulations performed.}
\label{fig.res_velxrms}
\end{figure}

%
Figure \ref{fig.res_velxrms} presents the contours of RMS of longitudinal velocity
fluctuations on a cut plane at $z/D=0$. One can notice the early development of the 
mixing layer when increasing the number of DOFs in the calculations. In Fig.\
\ref{fig.res_velxrmsc} the increase in the RMS values for longitudinal velocity 
fluctuation occurs right after the boundary condition. The same physical phenomenon 
occurs farther when decreasing the simulation resolution, which is noticeable when 
comparing Fig.s\ \ref{fig.res_velxrmsb} and \ref{fig.res_velxrmsa}. The contours of 
RMS of longitudinal velocity fluctuation also show that the fluctuation levels get 
smaller with earlier development of the mixing layer. In Fig.\ \ref{fig.res_velxrmsc}
the region of high velocity fluctuation is thinner and presents smaller values than in 
Fig.\ \ref{fig.res_velxrmsb}. In Fig.\ \ref{fig.res_velxrmsa}, the region with a high 
level of velocity fluctuation is the largest, and it presents the highest values when 
compared to the other two results.

\begin{figure}[htb!]
\centering
\subfloat[S1 simulation.]{
	\includegraphics[trim = 0mm 30mm 0mm 50mm, clip, width=0.80\linewidth]{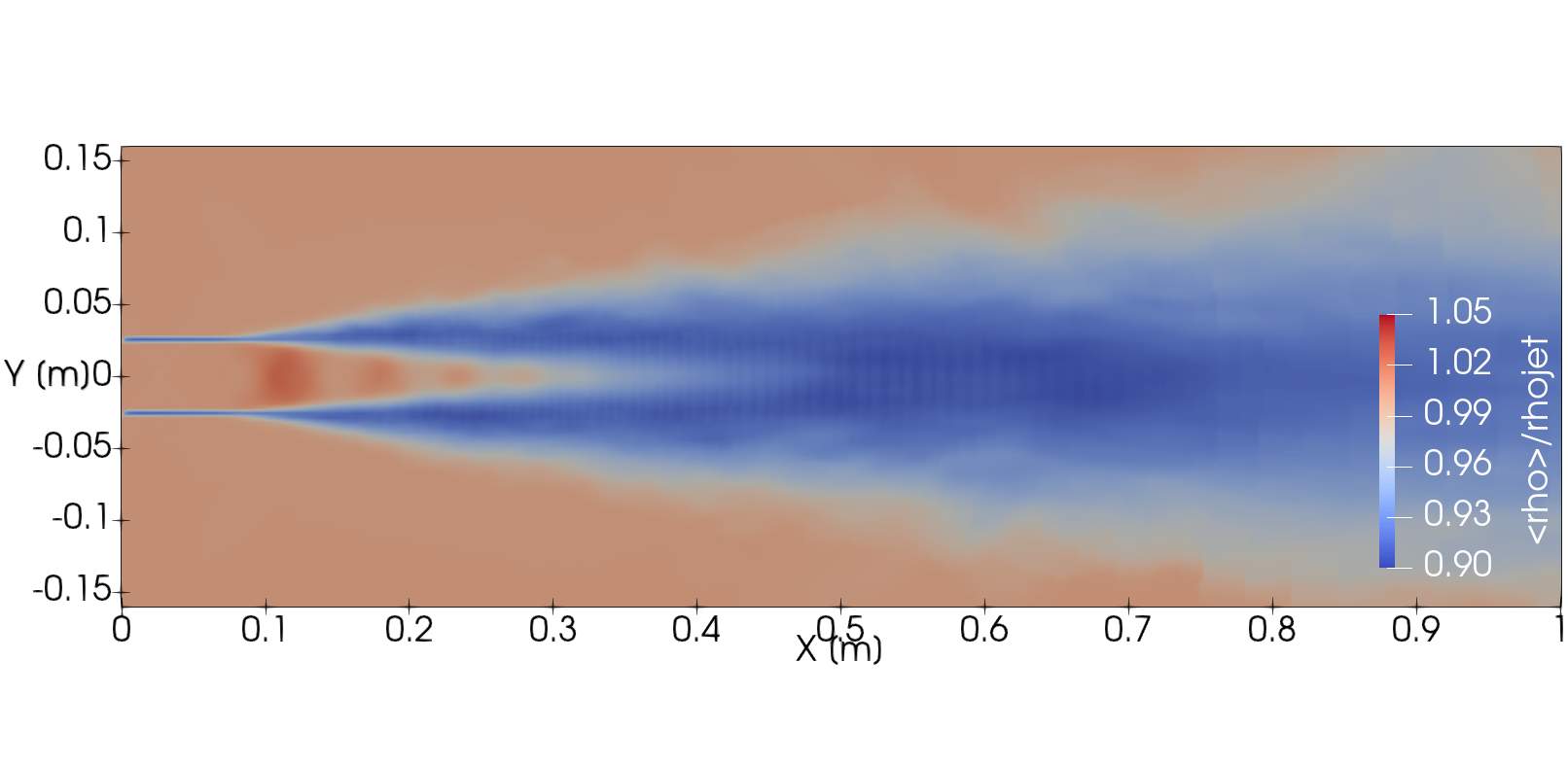}
	\label{fig.res_densa}
	}
\newline
\subfloat[S2 simulation.]{
	\includegraphics[trim = 0mm 30mm 0mm 50mm, clip, width=0.80\linewidth]{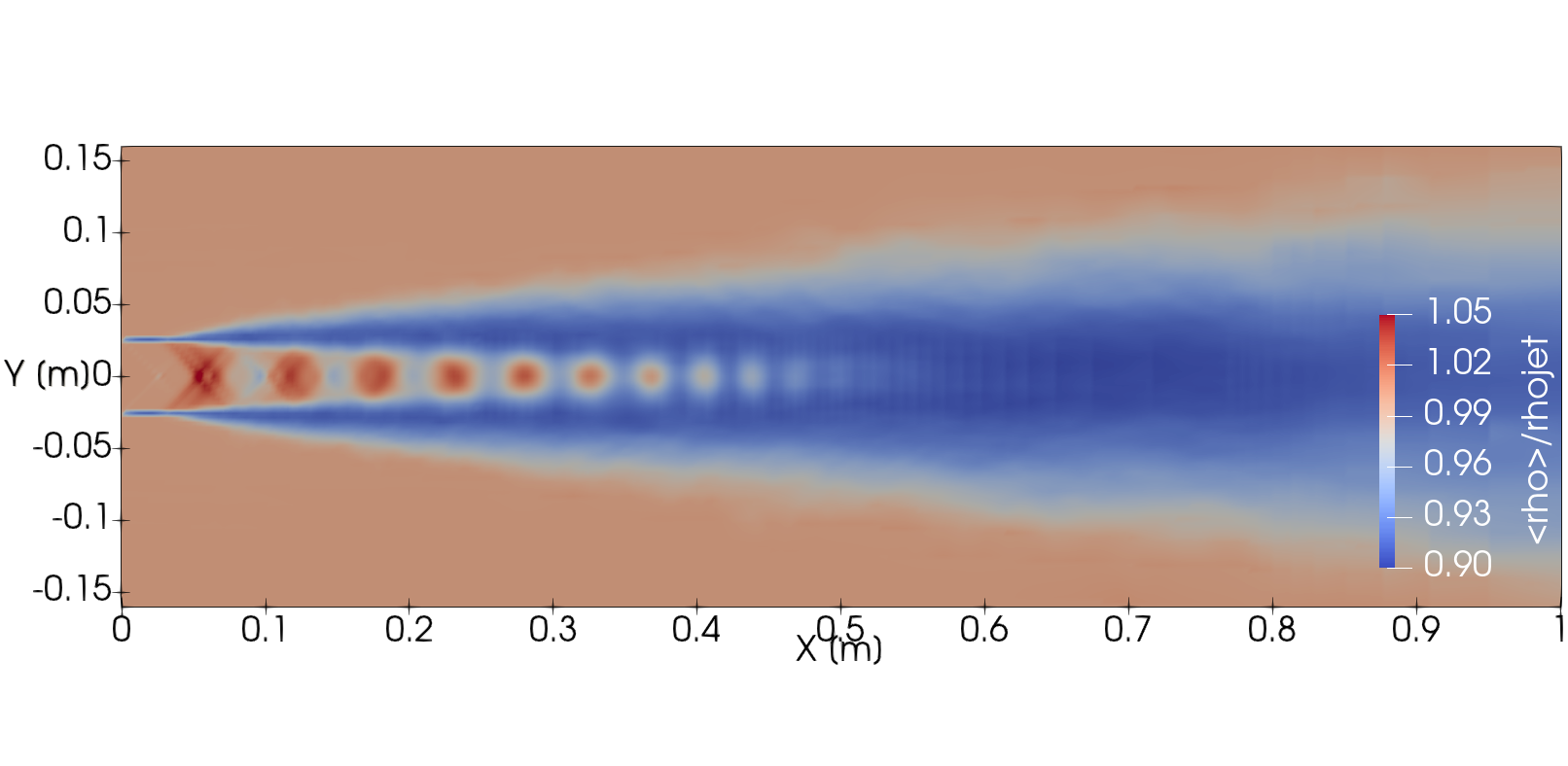}
	\label{fig.res_densb}
	}
\newline
\subfloat[S3 simulation.]{
	\includegraphics[trim = 0mm 30mm 0mm 50mm, clip, width=0.80\linewidth]{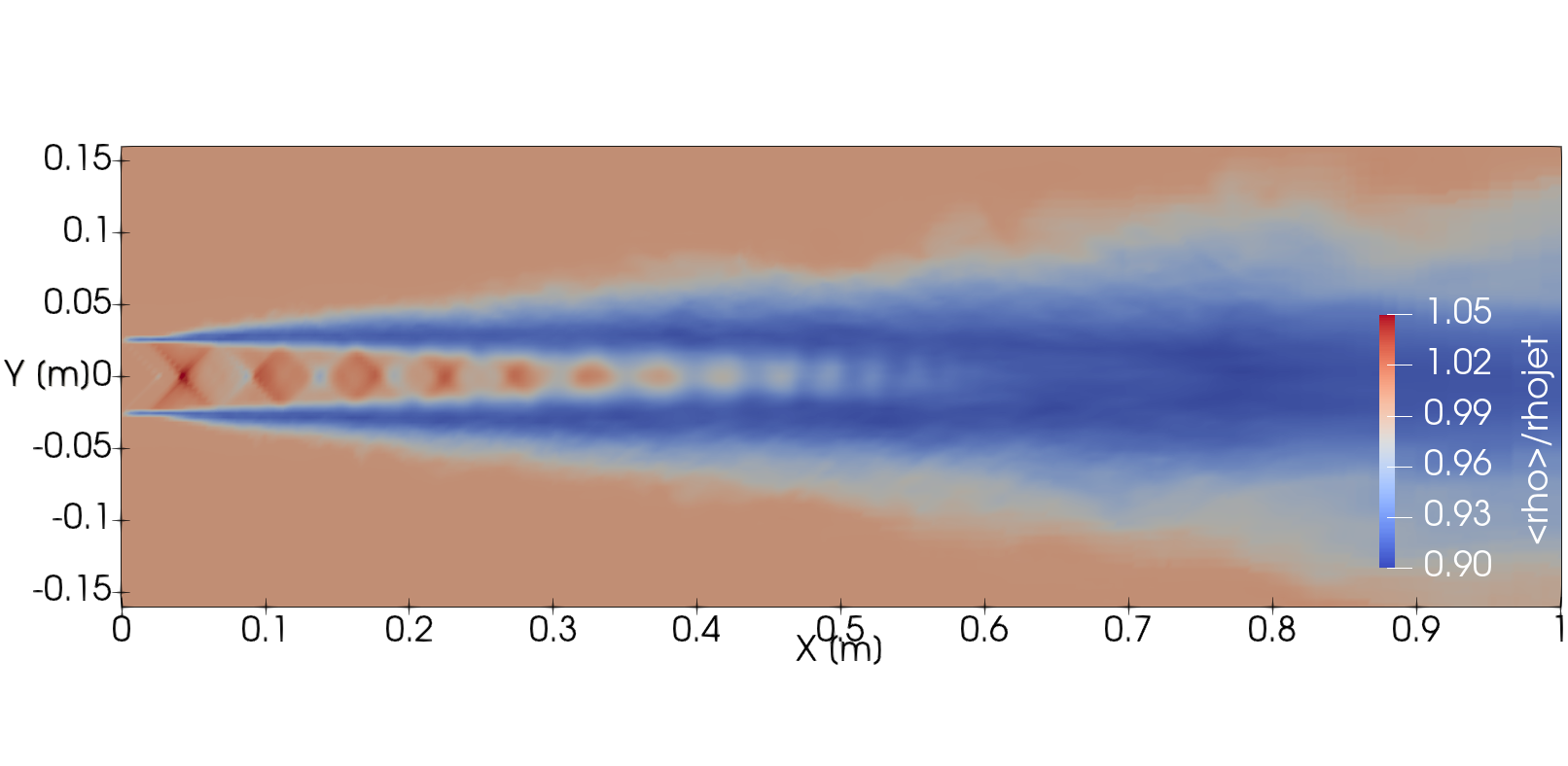}
	\label{fig.res_densc}
	}
\newline
\caption{Contours of mean density along cutplanes in $z/D=0$ for all three simulations.}
\label{fig.res_dens}
\end{figure}

The last contours presented in this section, Fig.\ \ref{fig.res_dens}, compare the mean
density results from the three simulations on a cut plane at $z/D=0$. It is possible 
to observe that each simulation presents different characteristics. In Fig.\
\ref{fig.res_densa} the shock waves are weak, being hardly visible with adequate range 
scales for all simulations. Moreover, one can notice a few shock waves and expansion 
waves reflections. The appearance of the first shock waves occurs far from the jet inlet
boundary condition. Figure \ref{fig.res_densb} presents a different behavior of the flow 
with shock waves and expansion waves significantly different from those observed in Fig.\
\ref{fig.res_densa}. The first shock waves appear closer to the jet inlet boundary
conditions, and they are visible, which indicates that they are significantly stronger 
than those from the S1 simulation. Another interesting observation is the number of 
shock waves and expansion waves reflection, which is much larger than the presented in 
Fig.\ \ref{fig.res_densa}. The final density results are presented in Fig.\
\ref{fig.res_densc}, which is the result from S3 simulation. It is possible to observe 
that the shock and expansion waves are better defined, presenting smaller thickness, 
which is also evidence of the improvements when increasing the number of DOFs in the
simulation. When comparing the results from Fig.\ \ref{fig.res_densc} with Fig.\
\ref{fig.res_densb}, it is possible to observe that the intensity of the shock waves 
is stronger in S2 simulation than in S3, even with the larger thickness. This is in 
agreement with results presented in Figs.\ \ref{fig.res_velxb} and \ref{fig.res_velxc}, 
in which the shock waves from S2 calculation where more visible. The quantity of shock 
waves reflections in the S2 and S3 test cases are similar.


\subsection{Velocity Profiles}

\begin{figure}[htb]
\centering
\subfloat[Centerline, mean.]{
	\includegraphics[width=0.47\linewidth]{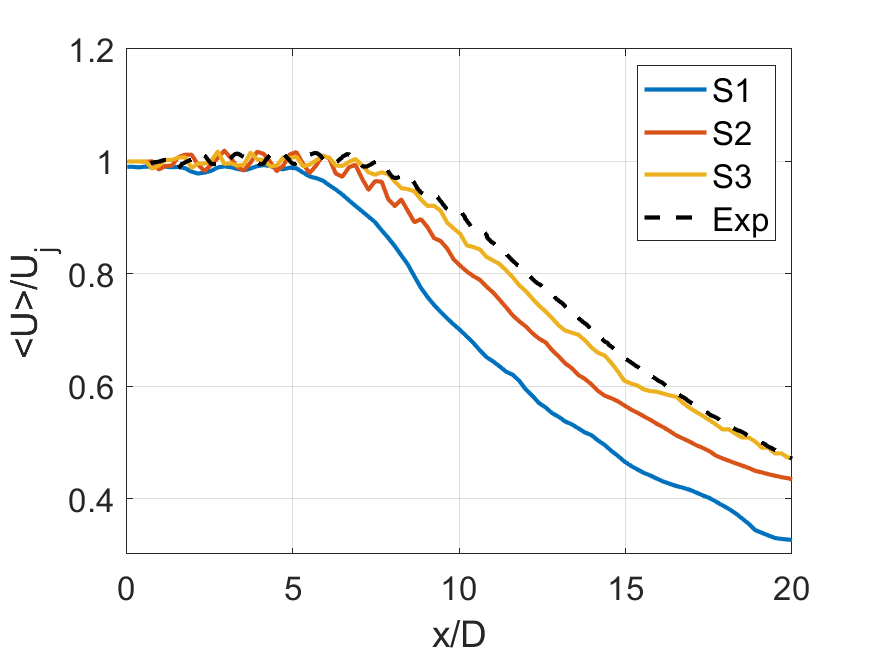}
	\label{fig.res1}	
	}%
\subfloat[Centerline, RMS values.]{
	\includegraphics[width=0.47\linewidth]{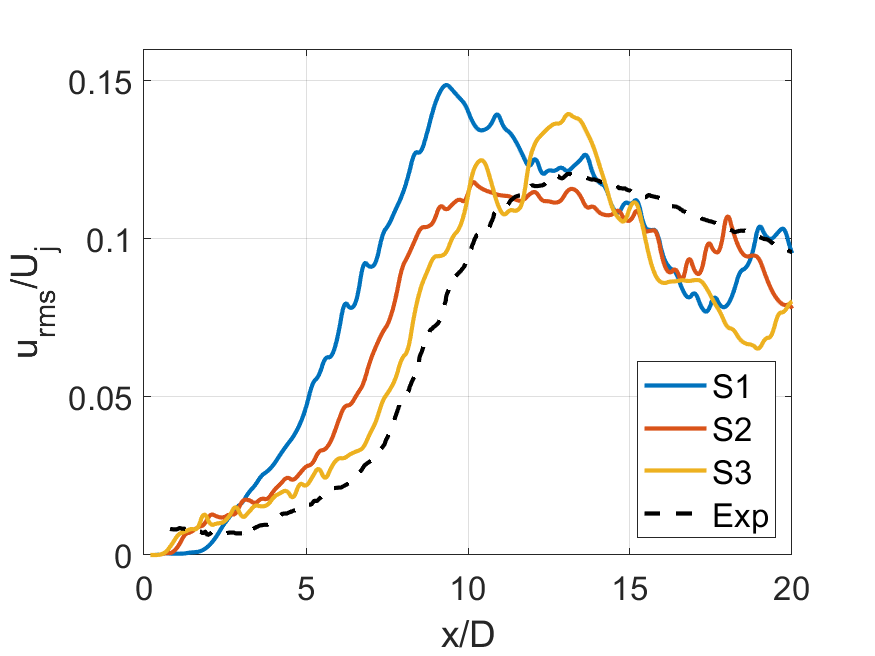}
	\label{fig.res2}	
	}
\newline
\subfloat[Lipline, mean.]{
	\includegraphics[width=0.47\linewidth]{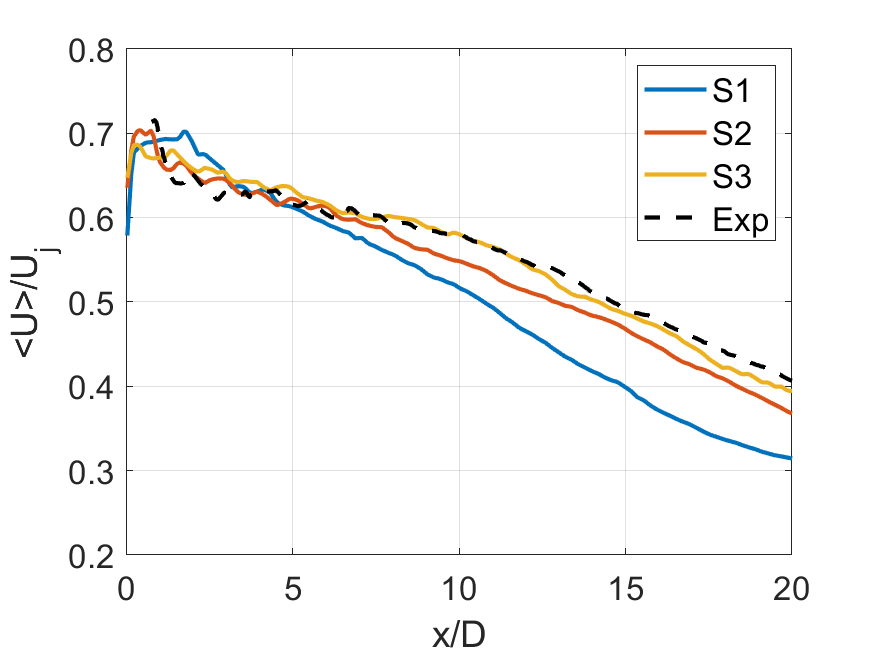}
	\label{fig.res3}
	}%
\subfloat[Lipline, RMS values.]{
	\includegraphics[width=0.47\linewidth]{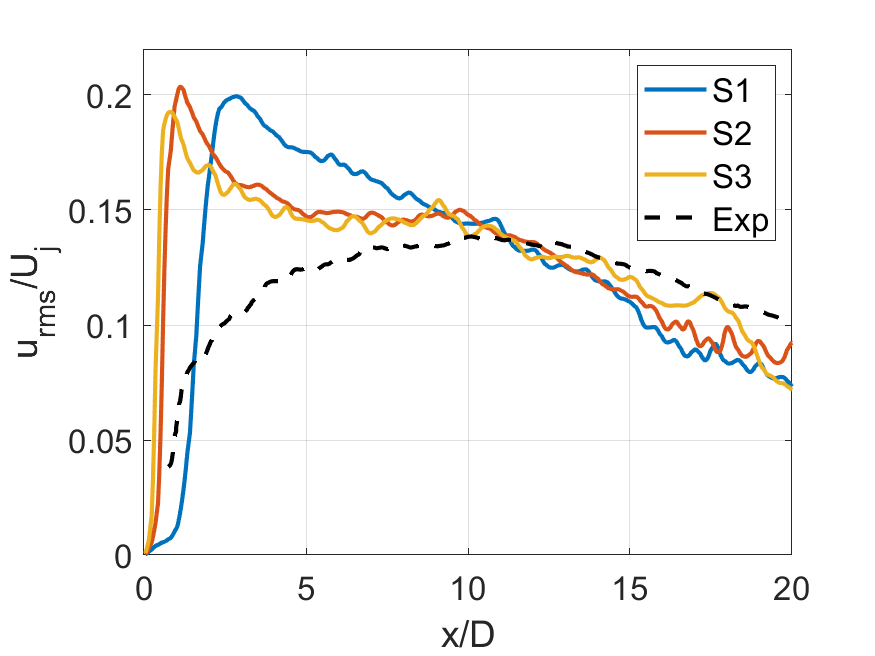}
	\label{fig.res4}	
	}
\caption{Results for the mean streamwise velocity component distributions (left) and RMS values of
streamwise velocity component fluctuations (right) in the jet centerline, $y/D=0$ (top), and lipline,
$y/D=0.5$ (bottom).}
\label{fig.res1_4}
\end{figure}

%
One can better understand the effects of the {\it hp} refinement on the numerical solution when
comparing the results to the experimental data. Figure \ref{fig.res1_4} presents the mean
longitudinal velocity component and the RMS values of the velocity fluctuation distributions
along the jet centerline ($y/D=0$) and the jet lipline ($y/D=0.5$). Analyzing the results
presented in Fig.\ \ref{fig.res1}, the improvement in the capacity to capture flow features
when increasing the numerical resolution of the calculations is prominent. One can notice 
the numerical solution progression of the mean longitudinal velocity component towards the
experimental data in Fig.\ \ref{fig.res1}. The potential core is longer in S3 than in the
other cases, which is presented in detail in Tab.\ \ref{tab.potcore}. Moreover, the most refined numerical study presents the intensity of the shock
waves and the slope of the decay of velocity comparable to the ones of the reference data.
One can observe that the profiles calculated in S2 and S3 simulation are closer than the
ones computed in S1 and S2, even with a higher ratio of degrees of freedom, 
$DOF_{S3}/DOF_{S2} \approx 3.42$ and $DOF_{S2}/DOF_{S1} \approx 2.4$. The slight improvement, 
even with a higher DOF ratio, can also indicate that the simulation S3 is very close to provide 
the converged solution for the chosen modeling approach. 
%

\begin{table}
\caption{Summary of potential core length for all simulations.}
\centering
\begin{tabular}{c c c} \hline \hline
Simulation     & Potential core & Error to  \\
  & length ($x/D$) & experimental data ($\%$) \\ \hline
 S1  & 6.3 & 30.0 \\
 S2  & 7.8 & 13.3\\
 S3  & 8.5 & 5.5\\ \hline \hline
\end{tabular}
\label{tab.potcore}
\end{table}

%
The results presented in Fig.\ \ref{fig.res2} agree well with the results of Fig.\ \ref{fig.res1}. The increase in the simulation resolution improves the calculation of RMS velocity fluctuation towards the experimental data. 
Analyzing the results close to the jet inlet, the increase in the velocity fluctuation
occurs further upstream in S3 simulation, which is in agreement with the contours in 
Figs.\ \ref{fig.res_velx} where the shock waves of S3 numerical study appear closer to jet 
inlet when compared to the S2 and S1 calculations. However, even with an early increase 
in the fluctuation levels in S3 computations, the slope of its profile in Fig.\ 
\ref{fig.res2} is smoother and closer to the reference than the one calculated in 
S2 and S1 numerical studies. In the same image, S3 simulation presents two peaks of velocity 
fluctuation, with the second one close to the peak indicated in the experiment. However, 
its RMS fluctuations are higher than experimental data and S2 simulation solution. 
The presence  of small values of velocity fluctuation close to 
the jet inlet could be related to imposed jet entrance boundary conditions.

%
Figures \ref{fig.res3} and \ref{fig.res4} illustrate the profiles of mean and RMS fluctuations of the 
longitudinal velocity component along the lipline, respectively. One can notice the improvements in the
simulation resolution with S3 and S2 simulations providing mean profiles closer to the experimental one than
the results from the S1 calculation. 
The mean velocity oscillations in the vicinity of the inlet jet may be correlated with 
shock waves. They are also present in the experimental profile along the lipline. The S1 and 
S2 simulations present an early reduction of mean velocity when comparing to the reference
profile. The most refined calculation has a mean velocity profile in good agreement with the
experimental data along the lipline. Such behavior is also noticeable along the centerline.
%
The calculations performed in the present paper have generated fluctuation profiles of the longitudinal
velocity fluctuation along the lipline, Fig.\ \ref{fig.res4}, that indicate a different trend from what
is stated in Figs.\ \ref{fig.res1} to \ref{fig.res3}. One can notice that, when increasing the simulation
resolution, the peak of velocity fluctuation moves towards the jet inlet, at $x/D=2.5$ for S1 simulation
and $\approx x/D=1$ for S2 and S3 calculations. This outcome is present in the contours indicated in 
Fig.\ \ref{fig.res_velx}, where the initial spreading of the mixing layer in S2 and S3 simulations starts
earlier than in the S1 calculation. Such behavior is different from the experimental RMS profiles along
the lipline in which the increase in the velocity fluctuation occurs slowly before reaching a plateau
between $x/D=5$ to $x/D=15$. Using a flat hat velocity profile at the jet entrance for the numerical
calculations can explain the divergence between the fluctuations obtained from the numerical approach and
the experiments near the inlet domain. Such boundary condition neglects the turbulent boundary layer 
effects carried from the nozzle to the jet flow.

\FloatBarrier

\begin{figure}[htb]
\centering
\subfloat[$x/D=2.5$]{
	\includegraphics[trim = 30mm 0mm 30mm 0mm, clip, width=0.2\linewidth]{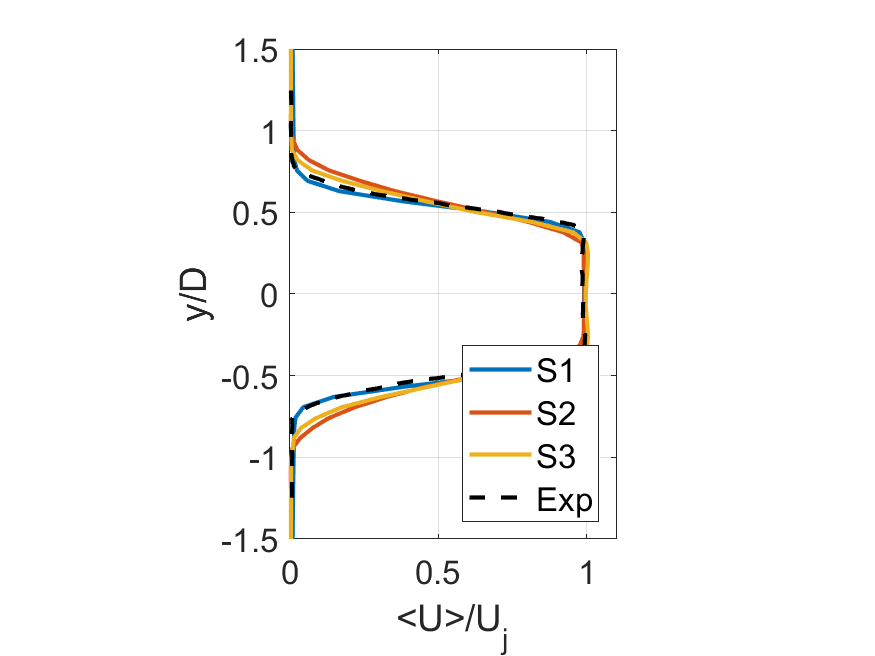}
	\label{fig.res5a}	
	}
\subfloat[$x/D=5$]{
	\includegraphics[trim = 30mm 0mm 30mm 0mm, clip, width=0.2\linewidth]{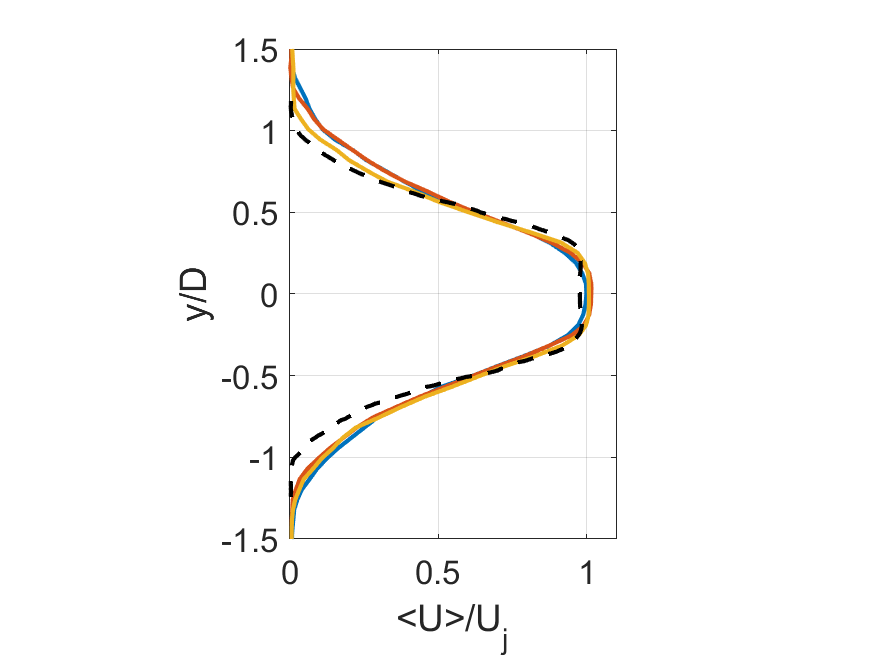}
	\label{fig.res5b}	
	}
\subfloat[$x/D=10$]{
	\includegraphics[trim = 30mm 0mm 30mm 0mm, clip, width=0.2\linewidth]{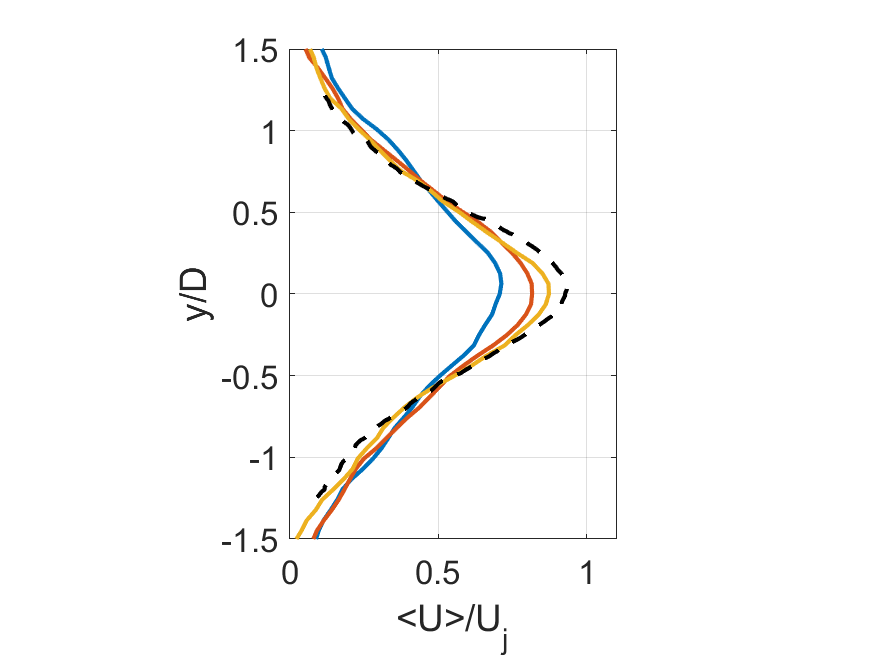}
	\label{fig.res5c}
	}
\subfloat[$x/D=15$]{
	\includegraphics[trim = 30mm 0mm 30mm 0mm, clip, width=0.2\linewidth]{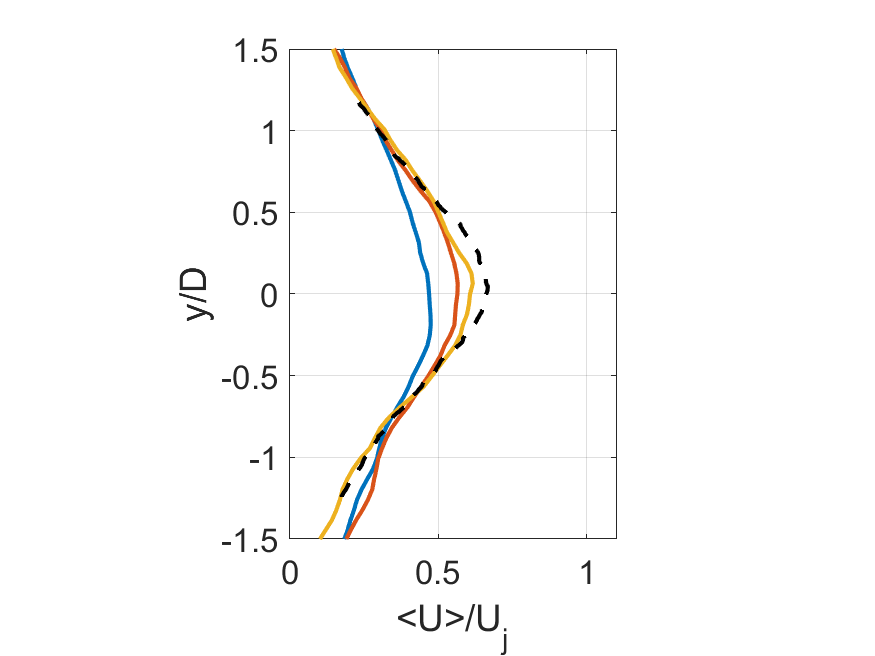}
	\label{fig.res5d}	
	}
\newline
\subfloat[$x/D=2.5$]{
	\includegraphics[trim = 30mm 0mm 30mm 0mm, clip, width=0.2\linewidth]{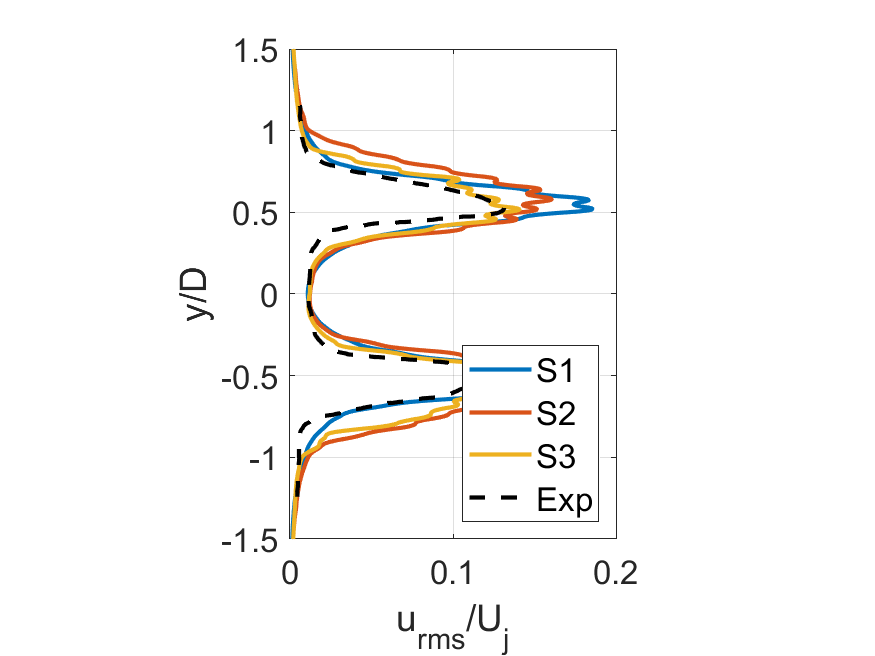}
	\label{fig.res5e}	
	}
\subfloat[$x/D=5$]{
	\includegraphics[trim = 30mm 0mm 30mm 0mm, clip, width=0.2\linewidth]{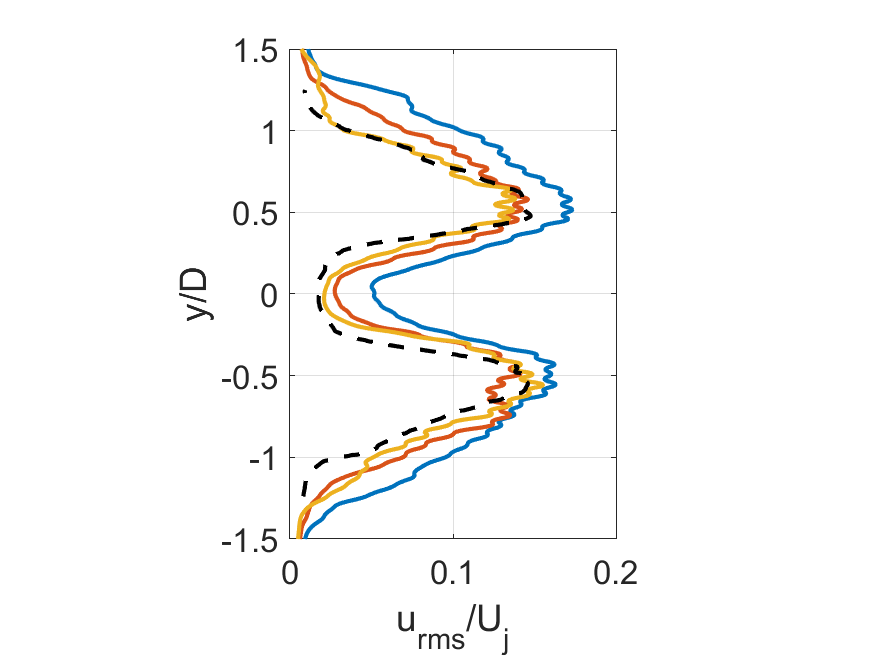}
	\label{fig.res5f}	
	}
\subfloat[$x/D=10$]{
	\includegraphics[trim = 30mm 0mm 30mm 0mm, clip, width=0.2\linewidth]{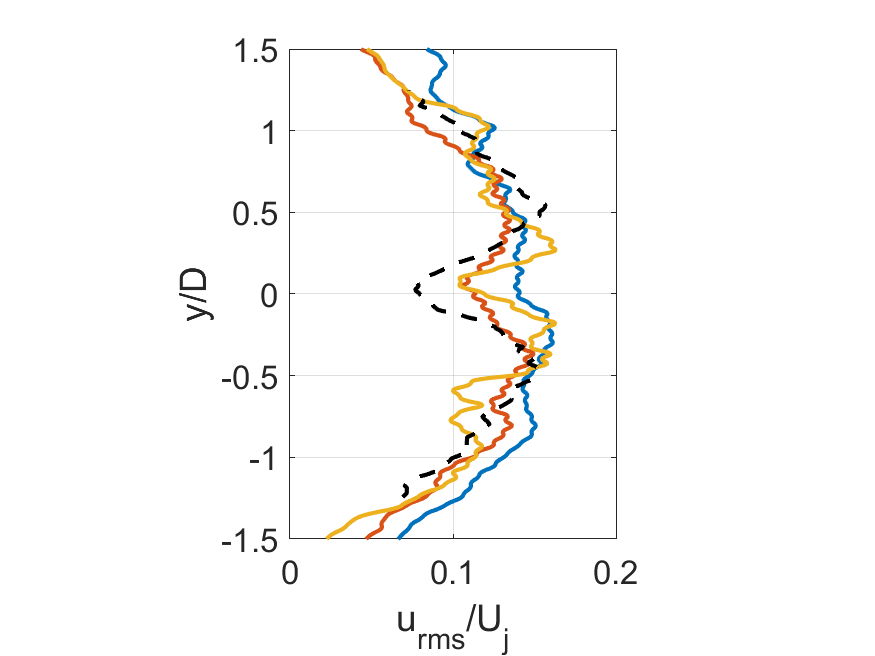}
	\label{fig.res5g}
	}
\subfloat[$x/D=15$]{
	\includegraphics[trim = 30mm 0mm 30mm 0mm, clip, width=0.2\linewidth]{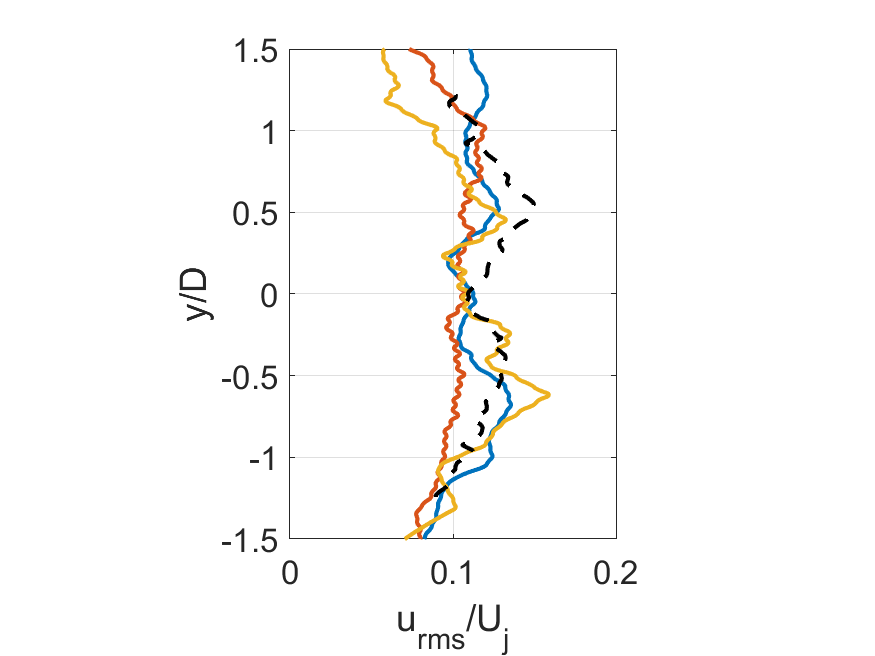}
	\label{fig.res5h}	
	}
\newline
\subfloat[$x/D=2.5$]{
	\includegraphics[trim = 30mm 0mm 30mm 0mm, clip, width=0.2\linewidth]{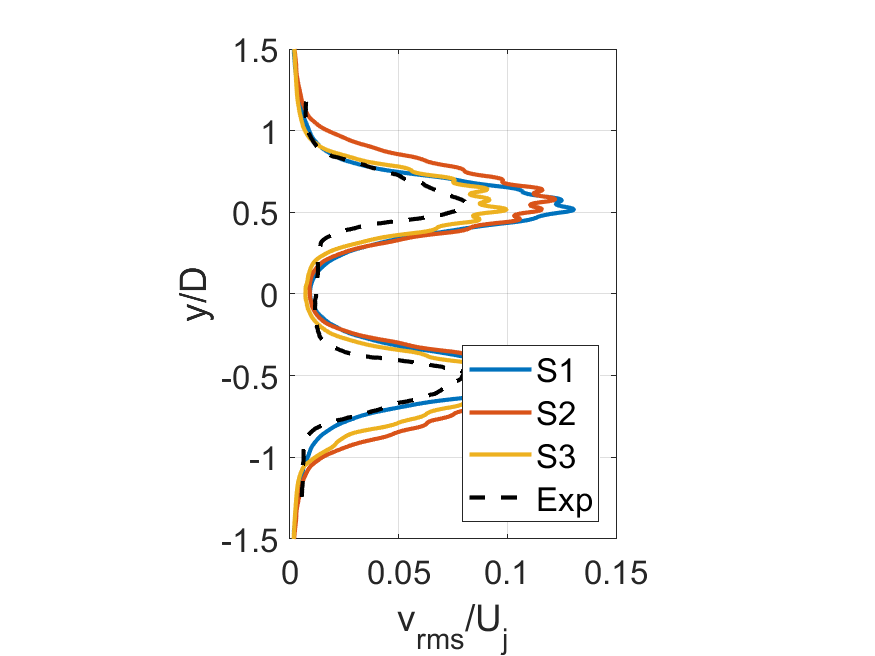}
	\label{fig.res5i}	
	}
\subfloat[$x/D=5$]{
	\includegraphics[trim = 30mm 0mm 30mm 0mm, clip, width=0.2\linewidth]{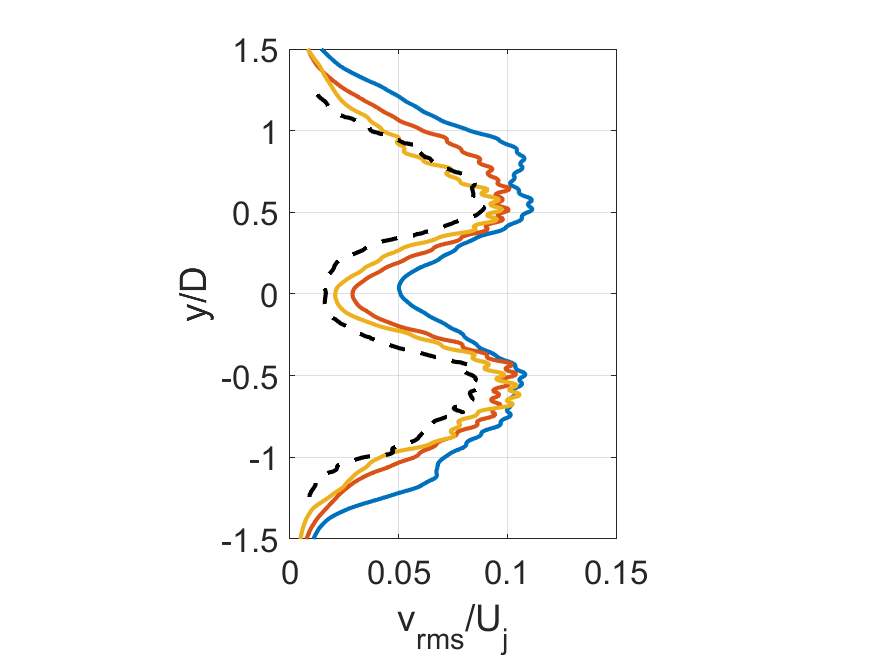}
	\label{fig.res5j}	
	}
\subfloat[$x/D=10$]{
	\includegraphics[trim = 30mm 0mm 30mm 0mm, clip, width=0.2\linewidth]{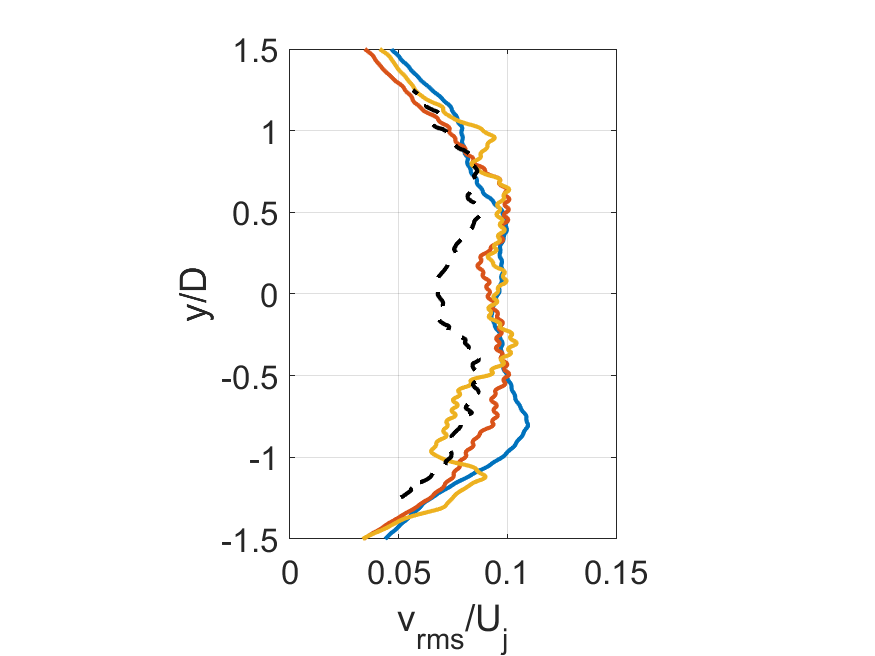}
	\label{fig.res5k}
	}
\subfloat[$x/D=15$]{
	\includegraphics[trim = 30mm 0mm 30mm 0mm, clip, width=0.2\linewidth]{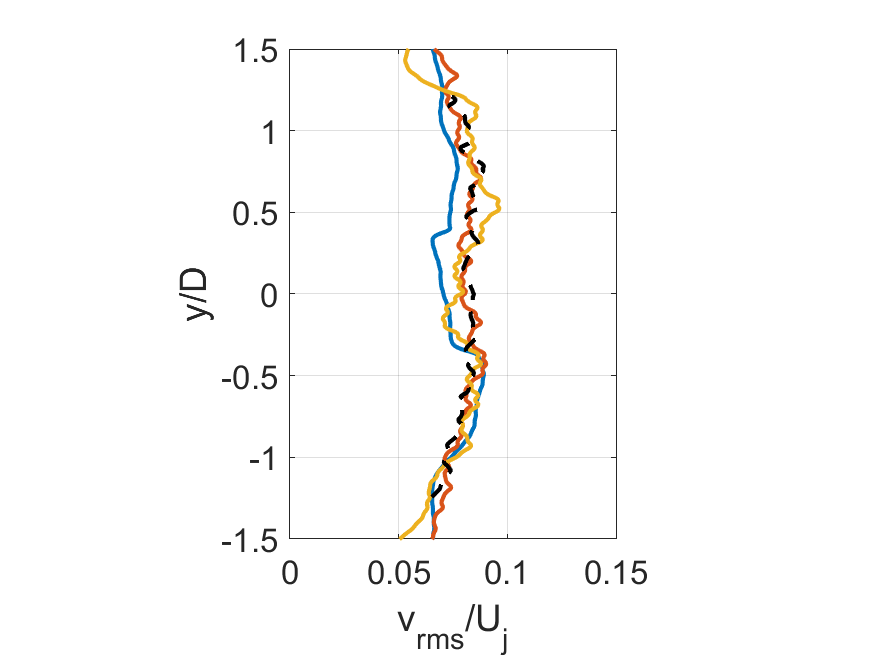}
	\label{fig.res5l}	
	}
\newline
\subfloat[$x/D=2.5$]{
	\includegraphics[trim = 30mm 0mm 30mm 0mm, clip, width=0.2\linewidth]{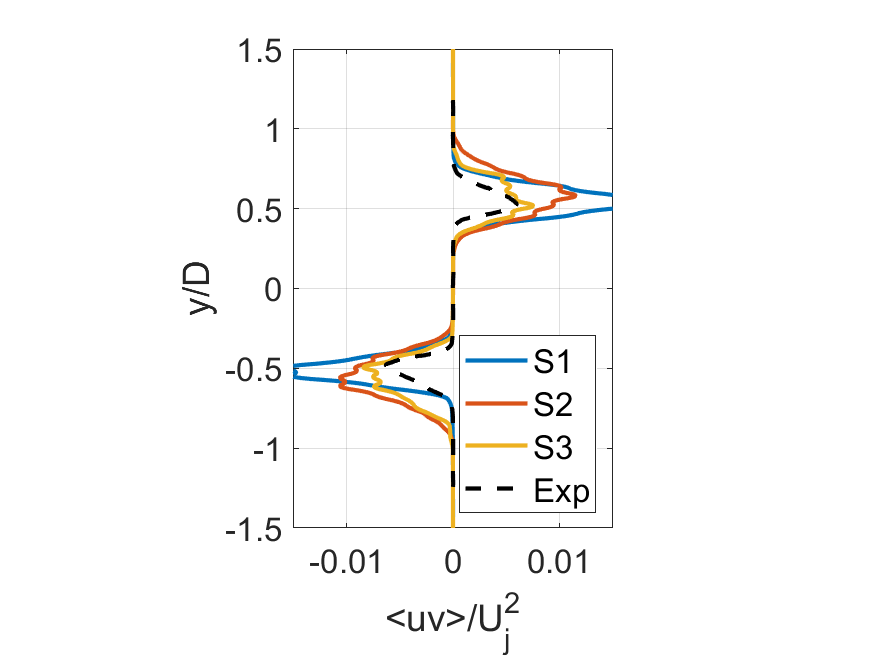}
	\label{fig.res5m}	
	}
\subfloat[$x/D=5$]{
	\includegraphics[trim = 30mm 0mm 30mm 0mm, clip, width=0.2\linewidth]{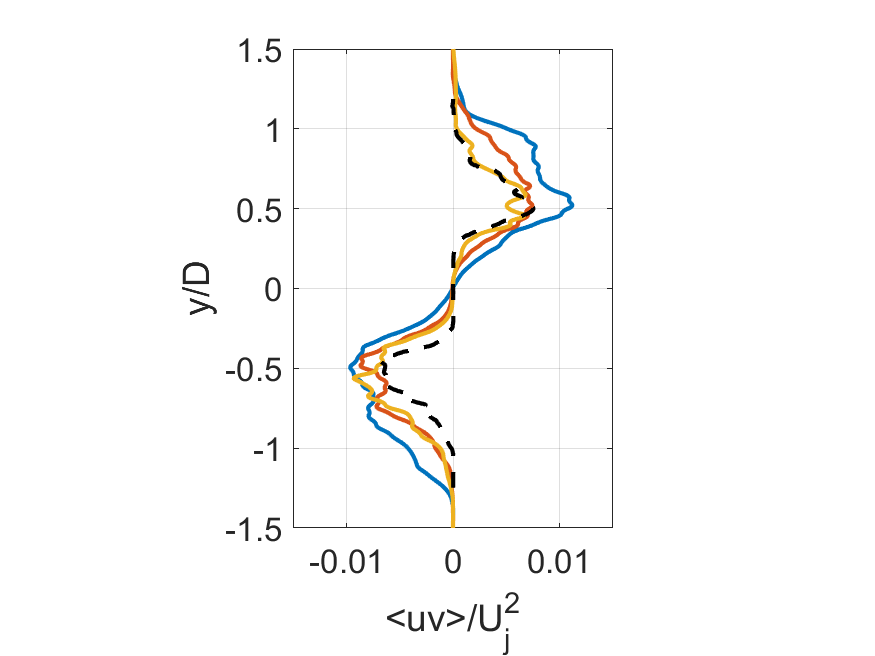}
	\label{fig.res5n}	
	}
\subfloat[$x/D=10$]{
	\includegraphics[trim = 30mm 0mm 30mm 0mm, clip, width=0.2\linewidth]{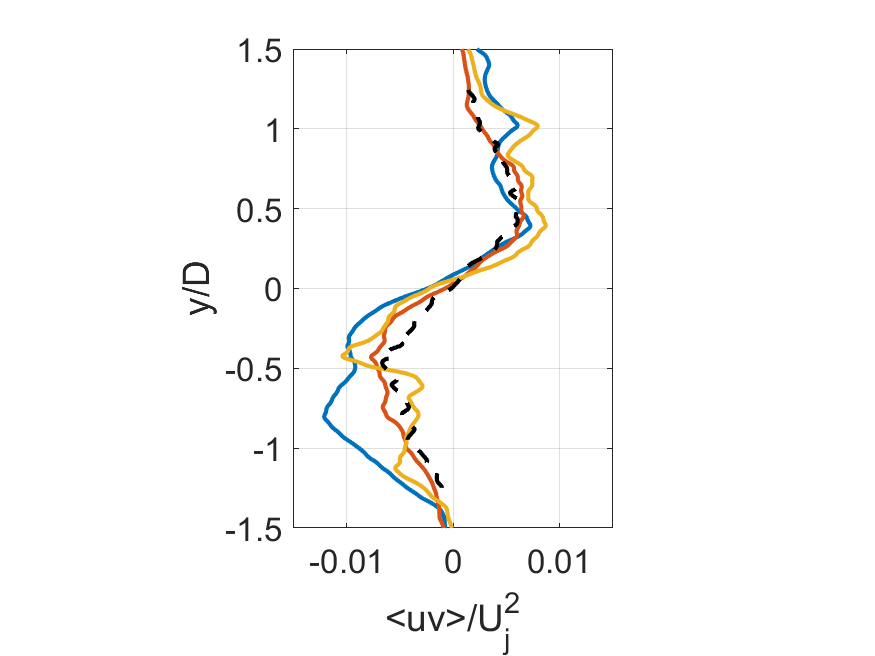}
	\label{fig.res5o}
	}
\subfloat[$x/D=15$]{
	\includegraphics[trim = 30mm 0mm 30mm 0mm, clip, width=0.2\linewidth]{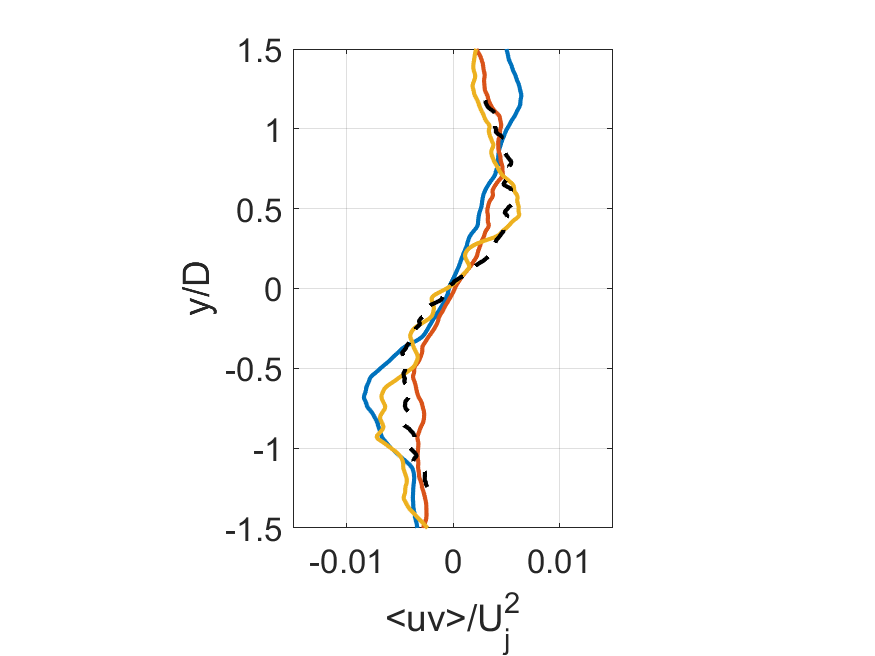}
	\label{fig.res5p}	
	}
\newline
\caption{Profiles of mean streamwise velocity component, RMS of streamwise velocity 
fluctuation, RMS of radial velocity fluctuation, and mean shear-stress tensor 
component (from top to bottom) at four streamwise positions $x/D=2.5$, $x/D=5$, $x/D=10$ and 
$x/D=15$ (from left to right).}
\label{fig.res5}
\end{figure}

%
Figure \ref{fig.res5} displays different statistical properties of the flow in four streamwise positions,
$x/D = 2.5$, $x/D=5$, $x/D=10$ and $x/D=15$. Figures \ref{fig.res5a} to \ref{fig.res5d} present the mean
longitudinal velocity component, Figs.\ \ref{fig.res5e} to \ref{fig.res5h} illustrate the RMS of the
longitudinal velocity fluctuation, Figs.\ \ref{fig.res5i} to \ref{fig.res5l} introduces the RMS of the
radial velocity fluctuation, and Figs.\ \ref{fig.res5m} to \ref{fig.res5p} indicate the mean shear stress
tensor.

%

The first set of results, in Figs.\ \ref{fig.res5a} to \ref{fig.res5d}, explicit some aspects of the
numerical results not investigated by the comparison of 2-D field of properties. In the first 
longitudinal position, Fig.\ \ref{fig.res5a}, S1 calculations generated mean profiles that are in 
better agreement with the experimental data than the ones from S2 and S3 numerical studies, which 
indicate a larger spreading of velocity at this position. The early development of the mixing layer 
from S2 and S3 simulations reinforces the influence of the choice of boundary conditions imposed. 
Analyzing the profiles from downstream positions, Figs.\ \ref{fig.res5b} to \ref{fig.res5d}, it is 
possible to verify the large spreading of velocity from the S1 simulation, with a reduction of 
longitudinal velocity in the jet centerline when compared to the other numerical solutions. 
Calculations with higher resolution can better capture the experimental trends, with the simulation 
S3 getting closer to experimental data.


%
The profiles of RMS values of streamwise velocity fluctuation are indicated in Figs.\ \ref{fig.res5e} to
\ref{fig.res5h}. The numerical results at $x/D=2.5$ present a similar profile to the one from the 
reference. However, the peaks generated by the calculations are higher than the experimental ones, 
with the results from the S3 simulation being the closest to experimental data. The same conclusion 
can be drawn for $x/D=5.0$, Fig.\ \ref{fig.res5f}.  In Figs.\ \ref{fig.res5g} and \ref{fig.res5h}, 
the all numerical results present a shape similar to experimental data, with a nearly constant value 
of velocity fluctuation. In Fig.\ \ref{fig.res5g} the experimental data still present a small level 
of fluctuation close to the centerline, which is not seen in the numerical profiles.

%
Profiles of RMS values of radial velocity component fluctuation are presented in Figs.\ \ref{fig.res5i} 
to \ref{fig.res5l}. In the first two longitudinal positions, Figs.\ \ref{fig.res5i} and \ref{fig.res5j},
the numerical results present a larger peak of fluctuation than the one from the experimental data, with
the profiles from the S3 simulation getting closer to reference. In Figs.\ \ref{fig.res5k} and
\ref{fig.res5l} the RMS profiles from the calculations are very similar, with the centerline of the
experimental data presenting small values of fluctuation in Fig.\ \ref{fig.res5k}.

%
Mean shear-stress tensor component profiles are presented in Figs.\ \ref{fig.res5m} to \ref{fig.res5p}. 
In the first two positions, $x/D=0.5$ and $x/D=2.5$, the peak of the shear-stress tensor from numerical 
calculations is larger than the experimental one. Moreover, the peak region is wider than the one
observed in the experiments. In Fig.\ \ref{fig.res5n} the peaks are still larger than those observed 
in the reference. The differences between the simulations are smaller and closer to experimental data.
However, the region of the peaks is still wider than the one visualized in the experimental data. In 
Figs.\ \ref{fig.res5o} and \ref{fig.res5p}, the differences between numerical results and the
experimental data are reduced, and once more, all profiles are nearly matching.

\subsection{Summary of Discussions}

%
Results from the S2 simulation present significant improvements when compared to the solution from 
the S1 calculation. The increase in polynomial order of accuracy in the S3 study brings the results to 
a good agreement with the reference experimental data. 
%
%
%
The most significant point that does not present improvements when increasing the discretization resolution is related to the mean and fluctuating longitudinal velocity close to jet lipline, Figs.\ \ref{fig.res3} and\ref{fig.res4}. Grid refinement yields a velocity fluctuation peak that occurs closer to the jet inlet when compared to the experimental data. This behavior may be related to the hypothesis used to impose the inlet boundary condition, which neglects the effects of the jet boundary layer and leads to a not realistic turbulence transition in the vicinity of the nozzle.

%
The current work provides intel on the resolution requirements to perform large-eddy simulations of
supersonic jet flows. The next step concerns the exploration of alternatives to improve the jet
simulation in the lipline close to the jet inlet condition. A solution to improve the simulation
in this region is a better characterization of the flow leaving the nozzle. The choice of a uniform
velocity profile is preliminary, and it does not represent the physics of the flow. The continuity of
the work will focus on options to produce an experiment-like condition in the jet inlet condition.

\FloatBarrier

\section{Concluding Remarks}

%
The current work assesses the effects of mesh and polynomial ({\em hp}) refinement using a nodal 
discontinuous Galerkin methodology to evaluate the resolution requirements for large eddy simulations 
of compressible jet flows. The problem of interest is a perfectly expanded supersonic jet flow with 
a Mach number of $1.4$ and Reynolds number based on the jet exit diameter of $1.58 \times 10^6$. 
Initially, a mesh with $6.2 \times 10^6$ elements is used with interpolation polynomials that yield 
second-order spatial accuracy, in order to produce a starting point for the comparisons. Such 
calculations use, therefore, the equivalent to approximately $50 \times 10^6$ degrees of freedom 
(DOFs)\@. Afterwards, a new mesh is developed with some topological improvements and additional 
refinement, leading to $15.4 \times 10^6$ elements. The new mesh is applied in simulations using 
second-order and third-order spatial accuracy, resulting in $120 \times 10^6$ and 
$410 \times 10^6$ DOFs, respectively. The results for the simulations are compared to experimental data.

%
The paper initially investigates the contours of mean velocity, mean pressure, and velocity 
fluctuations. The comparison indicates that the mesh/polynomial refinement improves the jet 
calculations by enhancing the prediction of the mixing layer and of the series of shock and 
expansion waves in the jet core. Therefore, as one should expect, more refined computations 
lead to an improved ability to predict flow features, as one can see in the present paper by 
the comparison of the numerical solutions and the experimental data. 
Therefore, it is correct to state that the present paper indicates mesh and discretization 
parameters for LES-based calculations, using a nodal discontinuous Galerkin formulation, that 
provide supersonic jet flow results in good agreement with experiments. 
%
%

One important aspect that becomes clear in the present calculation results is that the jet inlet 
boundary condition, used in the current work, has a significant impact on the ability of representing 
the very early stages of jet mixing. In particular, this observation becomes evident by looking at 
the behavior of RMS values of fluctuating properties near the jet exit, along the jet lipline. 
All three simulations have failed to capture the correct mixing behavior, as evidenced by the 
comparison with the experimental data. Moreover, the increased numerical resolution, although providing 
much better comparisons for the overall solution, does not improve the behavior of fluctuating 
properties near the jet exit. Hence, the continuation of the present effort will address possible 
improvements in the jet inlet boundary conditions.


\section*{Acknowledgments}

The authors acknowledge the support for the present research provided by Conselho Nacional de Desenvolvimento Cient\'{\i}fico e Tecnol\'{o}gico, CNPq, under the Research Grant No.\ 309985/2013-7\@. The work is also supported by the computational resources from the Center for Mathematical Sciences Applied to Industry, CeMEAI, funded by Funda\c{c}\~{a}o de Amparo \`{a} Pesquisa do Estado de S\~{a}o Paulo, FAPESP, under the Research Grant No.\ 2013/07375-0\@. The authors further acknowledge the National Laboratory for Scientific Computing (LNCC/MCTI, Brazil) for providing HPC resources of the SDumont supercomputer. This work was also granted access to the HPC resources of IDRIS under the allocation 2021-A0112A12067 made by GENCI. The first author acknowledges authorization by his employer, Embraer S.A., which has allowed his participation in the present research effort. The doctoral scholarship provide by FAPESP to the third author, under the Research Grant No.\ 2018/05524-1, is thankfully acknowledged. Additional support to the fourth author under the FAPESP Research Grant No.\ 2013/07375-0 is also gratefully acknowledged.

\bibliography{bibfile_paper}

\end{document}